\documentclass[aip, pof,graphicx]{revtex4-2}

\usepackage{graphicx}
\usepackage{amssymb}
\usepackage{dcolumn}
\usepackage{bm}
\usepackage[utf8]{inputenc}
\usepackage[T1]{fontenc}
\usepackage{mathptmx}
\usepackage{color}
\usepackage{longtable}
\usepackage{epstopdf}

\draft 

\begin{document}


\title[Collaborative behavior of intruders moving amid grains]{Collaborative behavior of intruders moving amid grains\\
	This article may be downloaded for personal use only. Any other use requires prior permission of the author and AIP Publishing. This article appeared in Phys. Fluids 34, 123306 (2022) and may be found at https://doi.org/10.1063/5.0124556.} 



\author{Douglas D. Carvalho}
\affiliation{School of Mechanical Engineering, UNICAMP - University of Campinas,\\
	Rua Mendeleyev, 200, Campinas, SP, Brazil
}

\author{Erick M. Franklin}%
 \email{erick.franklin@unicamp.br}
 \thanks{Corresponding author}
\affiliation{School of Mechanical Engineering, UNICAMP - University of Campinas,\\
	Rua Mendeleyev, 200, Campinas, SP, Brazil
}


\date{\today}

\begin{abstract}
We investigate the motion of groups of intruders in a two-dimensional granular system by using discrete numerical simulations. By imposing either a constant velocity or a thrusting force on larger disks (intruders) that move within smaller ones (grains), we obtained instantaneous positions and components of forces for each intruder and grain. We found that (i) intruders cooperate even when at relatively large distances from each other; (ii) the cooperative dynamics is the result of contact chains linking the intruders as well as compaction and expansion of the granular medium in front and behind, respectively, each intruder; (iii) the collaborative behavior depends on the initial arrangement of intruders; and (iv) for some initial arrangements, the same spatial configuration is eventually reached. Finally, we show the existence of an optimal distance for minimum drag for a given set of intruders, which can prove useful for devices stirring the ground or other granular surfaces.
\end{abstract}


\maketitle 


\section{\label{sec:Intro} INTRODUCTION}

From the fast penetration of solid bodies in the ground (such as asteroids colliding with moons and planets) to the slower motion of animals and machines within grains (worms, insects and plows, for example), the displacement of solid objects (intruders) in a granular medium is frequent on Earth and other celestial bodies. When the velocities involved are relatively small, the motion of grains is dominated by solid-solid friction and intermittent contact chains \cite{Kolb1, Tordesillas, Kozlowski, Carlevaro, Kozlowski2, Kozlowski3, Pugnaloni, Carvalho}, this regime being called quasistatic. For higher velocities, a dense regime appears, where grains flow in a more continuous manner, but still subject to friction and contact chains, in addition to collisions. For still higher velocities, a dilute regime appears, in which solid-solid collisions dominate \cite{Andreotti_6}.

When the intruder's displacement promotes the quasistatic motion of grains, contact chains appear and collapse successively, forming a time-varying contact network that percolates forces within the granular medium. This intermittent network implies history dependence in the stress distribution, with the occurrence of local anisotropy, local packing variations, and jamming and unjamming regions as the contact chains persist or fail, respectively \cite{Radjai1, Majmudar, Cates, Majmudar, Bi, Seguin2, Behringer_1, Featherstone}. On the intruder, it causes a strongly oscillating drag force \cite{Kolb1, Seguin1, Carvalho}. Therefore, the problem is intricate even in the case of a single intruder.

Given the problem complexity, most of previous studies were devoted to the motion of one intruder within a granular medium \cite{Albert, Albert2, Stone, Geng, Costantino, Kolb1, Seguin1, Kozlowski, Carlevaro, Pugnaloni, Carvalho}, in general using two-dimensional systems, and, although using relatively simple setups, they brought important insights into the problem. For example, Kolb et al. \cite{Kolb1} and Seguin et al. \cite{Seguin1} investigated experimentally the dynamics of a granular medium consisting of disks being displaced by an intruder. They found compaction and expansion in front and behind the intruder, respectively, that jamming can occur in compacted regions, and that the drag force on the intruder presents strong fluctuations as contact chains successively form and collapse. In addition, Seguin et al. \cite{Seguin1} computed the macroscopic friction coefficient $\mu$ and showed that it depends on the azimuthal direction (and, therefore, a nonlocal rheology seems necessary in continuum models). Kozlowski et al. \cite{Kozlowski} and Carlevaro et al. \cite{Carlevaro} investigated experimentally and numerically the motion of an intruder driven by a loaded spring within a bidimensional granular medium, and were particularly interested in the effects of the packing fraction and interparticle and basal frictions. They showed that in the presence of basal friction there are two regimes: an intermittent regime at low particle fractions, in which the intruder moves freely between clogging events, and a stick-slip regime, in which the intruder alternates fast slip events with creeping over long times. They showed also that in the absence of basal friction only the intermittent regime is observed, and that the intermittent to stick-slip transition is affected by the dynamic coefficient of basal friction while it is roughly independent of the static coefficient. Later, Pugnaloni et al. \cite{Pugnaloni} showed that the stick-slip dynamics depends only on the sizes involved, being independent of friction coefficients, and proposed a model for the energy released by the spring as a function of the packing fraction. Tripura et al. \cite{Tripura} studied numerically how a two-dimensional granular medium consisting of single and pairs of disks (dumbbells) behaves when displaced by a larger intruder (single disk). They found that the drag force on the intruder increases with the proportion of dumbbells in the system, that the additional resistance caused by dumbbells is negligible when the microscopic friction is set to zero, and that the stress propagated in front of the intruder increases with its diameter. The problem was inquired further by Kozlowski et al. \cite{Kozlowski2, Kozlowski3}, who measured the effects of grain angularity on the stress propagation and stick-slip dynamics, showing that angular grains resist to motion under lower packing fractions and have higher shear strengths. Recently \cite{Carvalho}, we investigated numerically the motion of an intruder within a two-dimensional granular medium using a setup similar to that of Ref. \cite{Seguin1}. Among other findings, we showed that the contact network can be divided into a network carrying strong forces from the intruder toward the walls (bearing network) and another one carrying weak forces (dissipative network), that the force network can reach regions far downstream of the intruder, and that grains within the bearing chains creep while the chains break. The latter result explains how the load chains break and allow the intruder to proceed with its motion.

There are fewer studies concerning groups of intruders, most of them for intruders moving vertically in a light granular medium \cite{Pacheco, Solano, Dhiman, Kawabata, Pravin, Espinosa} (those where the material density of grains is much lower than that of the intruder), corresponding thus to gravity-packed  systems. Pacheco-V\'azquez and Ruiz-Su\'arez \cite{Pacheco} investigated the sinking of sets of intruders impacting a low-density granular system, and showed the existence of a cooperative dynamics. For a pair of intruders initially side by side, they found that they first repel at the impact and afterward attract each other (in the horizontal plane, transverse to their motion). While they explained the initial repulsion by an increase in the granular pressure between the intruders and the attraction by a Bernoulli-like mechanism, Dhiman et al. \cite{Dhiman} proposed an explanation based on the formation and collapse of contact chains. For a number of intruders slightly larger  (five, for instance) placed initially side by side, Pacheco-V\'azquez and Ruiz-Su\'arez \cite{Pacheco} showed that they assume upward and downward convex configurations in succession, depending on the initial intruder-intruder separation (above a certain value they fall in parallel). They explained this behavior based on a sequential increase and decrease of the drag on the central intruders caused by the compaction and expansion of the bed, respectively. Finally, they showed that the intruders always finish horizontally aligned, irrespective of their number, initial configuration (vertical, horizontal or grouped distributions of intruders), sizes and densities, which they explained by the compaction-expansion mechanism.

Later, Solano-Altamirano et al. \cite{Solano} investigated the attractive and repulsive forces acting on a pair of intruders sinking in a granular bed (following impact). They found that the initial repulsion exists only when the separation between intruders is less than 6 grain diameters and attraction when the separation is less than 5-6 times the intruder diameter and sinking velocities higher than 1 m/s. They proposed that repulsion is due to granular jamming in the region between the intruders while attraction is caused by high velocities of grains in that region (Bernoulli-like effect). Merceron et al. \cite{Merceron} investigated a similar problem, but for pairs of intruders driven upwards in a confined granular system. They found that there is a characteristic separation between intruders for which the dynamics of grains in front of one intruder is affected by the other, and that this characteristic length is roughly independent of the intruder size. For a pair of side-by-side sinking intruders, Dhiman et al. \cite{Dhiman} found the existence of a separation for maximum attraction, and also an equilibrium separation for which neither attraction nor repulsion occur. Pravin et al. \cite{Pravin} showed that the work used in the vertical displacement of a pair of rods within a granular bed varies with their separation, a maximum value existing for a separation of 3 grain diameters. The work then decreases for larger separations until a distance of 11 grain diameters is reached (beyond this distance there exists a plateau). However, they also found that, for larger intruder sizes, the peak in work is reached at larger separations. In common, all those studies hint to attraction and repulsion mechanisms that favor collaborative behaviors, and optimal distances that reduce drag. 

Even though the cooperative behavior of intruders was evinced and its mechanisms explained in previous works, many aspects remain open, such as the forced motion of intruders in the horizontal direction, and the roles of friction, mean packing fraction and contact chains on the overall dynamics. In this paper, we investigate numerically how a group of intruders interact with each other while moving horizontally in a two-dimensional granular system. The numerical setup consists of two or three larger disks (intruders) driven either at constant speed or thrusting force within an assembly of smaller disks (grains) confined in a rectangular cell. As in Ref. \cite{Carvalho}, the disks are 3D objects (low height cylinders) sliding over a flat surface with both static and dynamic coefficients of friction, and we made use of the open-source code LIGGGHTS \cite{Kloss, Berger} with the DESIgn toolbox \cite{Herman} to carry out DEM (discrete element method) computations. We show that (i) intruders cooperate even when at relatively large distances from each other; (ii) the cooperative dynamics is the result of contact chains linking the intruders as well as compaction and expansion of the granular medium in front and behind, respectively, each intruder; (iii) the collaborative behavior depends on the initial arrangement of intruders; and (iv) for some initial arrangements, the same spatial configuration is eventually reached. We propose a chart for the collaborative patterns, and we show the existence of an optimal distance for minimum drag for a given set of intruders, which can prove useful for devices stirring the ground or other granular surfaces.

In the following, Sec. \ref{sec:model} presents the model equations, Sec. \ref{sec:setup} the numerical setup, and Sec. \ref{sec:Res} the results for intruders driven either at constant velocity or thrusting force. Finally, Sec. \ref{sec:Conclu} presents the conclusions.

\section{\label{sec:model} MODEL DESCRIPTION}

We carried out {DEM} \cite{Cundall} simulations using the open-source code LIGGGHTS \cite{Kloss, Berger} with the DESIgn toolbox \cite{Herman}. The simulations consisted basically in computing the linear (Eq. (\ref{Fp})) and angular (Eq. (\ref{Tp})) momentum equations for disks experiencing friction with the bottom and lateral walls and between them (the top wall was absent). For that, we implemented the static and dynamic frictions between the disks and the bottom wall (basal friction) in the DESIgn toolbox, which we modeled as in Carlevaro et al. \cite{Carlevaro} and describe in Carvalho et al. \cite{Carvalho}. A detailed explanation of the model can be found in Ref. \cite{Carvalho}; we provide only the essentials here.

\begin{equation}
	m\frac{d\vec{u}}{dt}= \vec{F}_{c} + m\vec{g}
	\label{Fp}
\end{equation}

\begin{equation}
	I\frac{d\vec{\omega}}{dt}=\vec{T}_{c}
	\label{Tp}
\end{equation}

In Eqs. (\ref{Fp}) and (\ref{Tp}), $\vec{g}$ is the acceleration of gravity and, for each disk, $m$ is the mass, $\vec{u}$ is the velocity, $I$ is the moment of inertia, $\vec{\omega}$ is the angular velocity, $\vec{F}_{c}$ is the resultant of contact forces, given by Eq. (\ref{Fc}), and $\vec{T}_{c}$ is the resultant of contact torques, given by Eq. (\ref{Tc}).

\begin{equation}
	\vec{F}_{c} = \sum_{i\neq j}^{N_c} \left(\vec{F}_{c,ij} \right) + \sum_{i}^{N_w} \left( \vec{F}_{c,iw} \right)
	\label{Fc}
\end{equation}

\begin{equation}
	\vec{T}_{c} = \sum_{i\neq j}^{N_c} \vec{T}_{c,ij} + \sum_{i}^{N_w} \vec{T}_{c,iw}
	\label{Tc}
\end{equation}

In Eqs. (\ref{Fc}) and (\ref{Tc}), $\vec{F}_{c,ij}$ and $\vec{F}_{c,iw}$ are the contact forces between disks $i$ and $j$ and between disk $i$ and the wall, respectively, with the basal friction included in $\vec{F}_{c,iw}$, $\vec{T}_{c,ij}$ is the torque due to the tangential component of the contact force between disks $i$ and $j$, $\vec{T}_{c,iw}$ is the torque due to the tangential component of the contact force between disk $i$ and the vertical wall, $N_c$ - 1 is the number of disks in contact with disk $i$, and $N_w$ the number of disks in contact with the wall. Finally, the elastic Hertz-Mindlin contact model \cite{direnzo} is used in both $\vec{F}_{c,ij}$ and $\vec{F}_{c,iw}$.

\section{\label{sec:setup} NUMERICAL SETUP}

The numerical setup is similar to that of Ref. \cite{Carvalho}, excepting the size of the domain, the number of intruders, and the type of external forcing. The system consisted of an assembly of 3D disks (grains) over a horizontal wall, confined by vertical walls, and a set of two (duo) or three (trio) larger 3D disks (intruders) in different initial configurations. Both the grains and the intruders had one of their flat surfaces facing the bottom wall, and the intruders moved at either constant velocity or thrusting force among the grains. The dimensions and properties of each grain and intruder are the same as in Refs. \cite{Seguin1, Carvalho}: the intruders were of steel with diameter $d_{int} = 16$ mm and height $h_{int} = 3.6$ mm, and the grains were of polyurethane (PSM-4) with height $h_{g} = 3.2$ mm and two different diameters, $d_{s} = 4$ mm and $d_{l} = 5$ mm, in order to prevent crystallization \cite{Speedy}. The areas occupied by the small and large grains were roughly the same by assuring a ratio $N_{l}/N_{s} \approx 0.64$ between the numbers of small ($N_{s}$) and large ($N_{l}$) disks.

\begin{figure}[h!]
	\begin{center}
		\includegraphics[width=0.95\columnwidth]{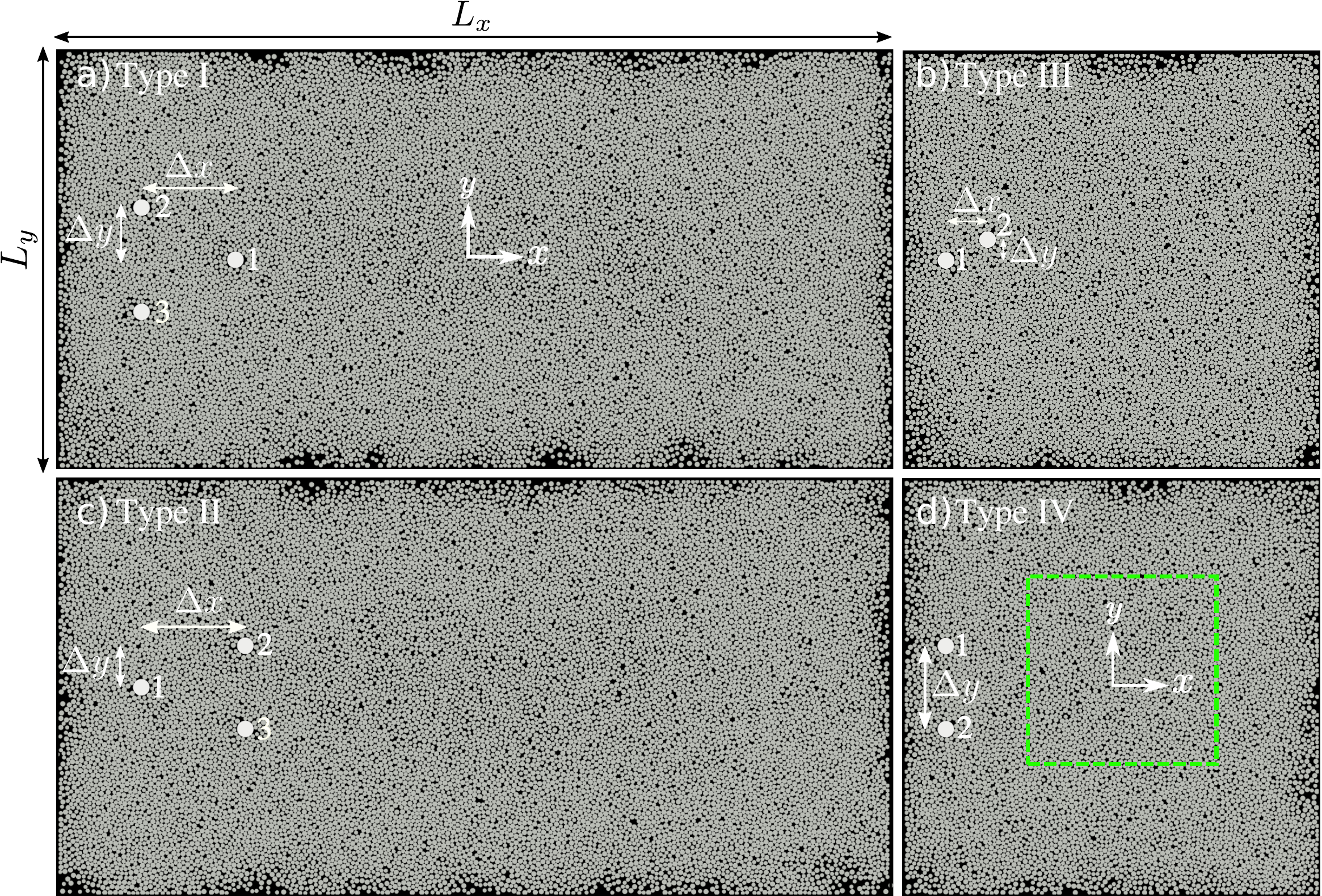}\\ 
	\end{center}
	\caption{Top-view images of the numerical setup for duos (d) aligned and (b) off-centered in the transverse direction, and trios with one intruder (a) in front and (c) behind two intruders initially aligned in the transverse direction. Figures (a) to (d) correspond to types I, III, II and IV, respectively. $\Delta x$ and $\Delta y$ are the initial separations in the longitudinal and transverse directions, respectively, and the area delimited by dashed-green lines in figure (d) is a static region of interest (ROI) used for some computations.}
	\label{fig:setup}
\end{figure}

The total domain varied depending on the simulation, being $L_x$ = $L_y$ = 0.4 m (for most simulations with two intruders) or $L_x$ = 0.8 and $L_y$ = 0.4 m (for all simulations with three and one with two intruders), where $L_x$ and $L_y$ are the longitudinal and transverse lengths, respectively. The domain was fixed for each simulation, so that the mean packing fraction $\phi$ remained constant in each run, but varied within 0.70 $\leq$ $\phi$ $\leq$ 0.79 for different runs by changing the number of disks in each simulation (the number of disks and the corresponding packing fractions are available in the supplementary material). We imposed the intruders to move at either a constant velocity of 2.7 mm/s (in cases with two intruders) or a constant external force (thrust) of 0.8 N (in cases with two and three intruders), and the motions were limited to the $xy$ plane, $x$ being the longitudinal and $y$ the transverse direction. We made use of 2.7 mm/s to allow for comparisons with the experiments of Seguin et al. \cite{Seguin1} (as done in Ref. \cite{Carvalho}), and 0.8 N to displace the intruders in the quasistatic regime (values slightly lower or higher would work as well). The 0.8 N thrust was imposed on each intruder, at each time step, in the $x$ direction. Examples of setups can be seen in Fig. \ref{fig:setup}.

The intruders were placed initially in the left region of the domain, by either duos or trios with initial separations $\Delta x$ and $\Delta y$ in the longitudinal and transverse directions, respectively, and were afterward put into motion toward the right region. We varied the initial values of $\Delta x$ and $\Delta y$ for different simulations, and, as the intruders were driven amid the grains, $\Delta x$ and $\Delta y$ were free to change along time in the cases of imposed thrust. Basically, four configurations were used: (i) one intruder in front of two intruders initially aligned in the transverse direction (Fig. \ref{fig:setup}(a)); (ii) one intruder behind two intruders initially aligned in the transverse direction (Fig. \ref{fig:setup}(c)); (iii) two intruders off-centered in the transverse direction (Fig. \ref{fig:setup}(b)); and (iv) two intruders initially aligned in the transverse direction (Fig. \ref{fig:setup}(d)). We call these configurations \textit{types I to IV}, respectively. The simulations with imposed velocity were carried out only for type IV, for different initial $\Delta y$, but in these cases, due to the forcing characteristics, $\Delta y$ remained constant along the time. Figure \ref{fig:setup}(d) also shows a region of interest (ROI) delimited by dashed-green lines (fixed in space), measuring 160 mm $\times$ 160 mm (corresponding to approximately 35$d_g$ $\times$ 35$d_g$, where $d_g$ = 4.5 mm is the average diameter of disks) and used for computing the time-average forces on the intruders and anisotropy within the granular system. 

Values of the coefficients of restitution $\epsilon$ and friction $\mu$ (both static and dynamic), Young's modulus $E$ and Poisson ratio $\nu$ were obtained from the literature \cite{Carlevaro, Hashemnia, Gondret, Zaikin}, and sensitivity tests varying those coefficients are available in Carvalho et al. \cite{Carvalho}. We note that we used a value of $E$ for the steel two orders of magnitude smaller than the real one ($E$ $=$ 1.96 $\times$ 10$^{11}$ Pa) in order to increase the time step while keeping a reasonable accuracy in the results \cite{Lommen}. We implemented the basal friction, both static and dynamic, by defining a threshold velocity $v'$ = 5 $\times$ 10$^{-4}$ m/s for the transition between static and dynamic conditions. Sensitivity tests for $v'$ are available in Carvalho et al. \cite{Carvalho}, where we verified that the time-averaged drag force does not change considerably for $v'$ $<$ 10$^{-3}$ m/s. Table \ref{tabmaterials} summarizes the mechanical properties of objects as used in the simulations and Tab. \ref{tabcoefficients} the values of friction and restitution coefficients.

\begin{table}[!h]
	\centering
	\caption{Properties of materials as used in the simulations: $E$ is the Young's modulus, $\nu$ is the Poisson ratio, $\rho$ is the material density, and $d$ is the particle diameter.}
	\label{tabmaterials}
	\begin{tabular}{l|c|c|c|c|c}
		\hline
		& \textbf{Material} & \textbf{$E$ (Pa)} & \textbf{$\nu$} & \textbf{$\rho$ (kg/m$^{3}$)}& \textbf{$d$ (mm)}\\
		\hline
		Intruder & Steel\footnotesize{$^{(1)}$} & $1.96 \times 10^{9}$  & 0.29 & 7800 & $d_{int} =$ 16            \\
		Grains & Polyurethane\footnotesize{$^{(1),(2)}$} & $4.14 \times 10^{6}$   & 0.50 & 1280 & $d_{s}$ = 4; $d_{l}$ = 5           \\
		Walls & Glass\footnotesize{$^{(1)}$} & $0.64 \times 10^{11}$ & 0.23 & 2500 & $\cdots$\\    
		\hline
		\multicolumn{3}{l}{\footnotesize{$^{(1)}$ Hashemnia and Spelt \cite{Hashemnia}}}\\
		\multicolumn{3}{l}{\footnotesize{$^{(2)}$ Gloss \cite{Gloss}}}
	\end{tabular}
\end{table}

\begin{table}[!h]
	\centering
	\caption{Friction and restitution coefficients used in the numerical simulations.}
	\label{tabcoefficients}
	\begin{tabular}{l|c|c}
		\hline
		\textbf{Coefficient}  & \textbf{Symbol} & \textbf{Value} \\
		\hline		  
		Restitution coefficient (grain-grain) & $\epsilon_{gg}$ & 0.30 \\
		Restitution coefficient (grain-intruder)\footnotesize{$^{(2)}$} & $\epsilon_{gi}$ & 0.70 \\
		Restitution coefficient (grain-wall)\footnotesize{$^{(3)}$} & $\epsilon_{gw}$ & 0.70 \\
		Restitution coefficient (intruder-intruder)\footnotesize{$^{(4)}$} & $\epsilon_{ii}$ & 0.56 \\
		Dynamic friction coefficient (grain-grain)\footnotesize{$^{(1)}$} & $\mu_{gg}$ & 1.20 \\
		Dynamic friction coefficient (grain-intruder)\footnotesize{$^{(2)}$} & $\mu_{gi}$ & 1.80 \\
		Dynamic friction coefficient (intruder-bottom wall) & $\mu_{iw}$ & 0.70 \\  
		Dynamic friction coefficient (grain-walls)\footnotesize{$^{(1)}$} & $\mu_{gw}$ & 0.40 \\
		Dynamic friction coefficient (intruder-intruder) & $\mu_{ii}$ & 0.57 \\		
		Static friction coefficient (grain-bottom wall) & $\mu_{s,gw}$ & 0.70 \\
		\hline
		\multicolumn{3}{l}{\footnotesize{$^{(1)}$ Carlevaro et al. \cite{Carlevaro}}} \\
		\multicolumn{3}{l}{\footnotesize{$^{(2)}$ Hashemnia et al. \cite{Hashemnia}}} \\
		\multicolumn{3}{l}{\footnotesize{$^{(3)}$ Gondret et al. \cite{Gondret}}}\\
		\multicolumn{3}{l}{\footnotesize{$^{(4)}$ Zaikin et al. \cite{Zaikin}}}
	\end{tabular}
\end{table}

Prior to each run, the grains were randomly distributed over a space larger than the computational domain, and then compressed toward the center until filling the desired domain. Afterward, the grains were allowed to relax and the simulation started. This assured the desired packing fraction within reasonable times (computation times for placing random grains directly under high packing fractions are prohibitive). The simulations were carried out with a time step $\Delta t = 3.2 \times 10^{-6}$ s, which was less than 10 \% of the Rayleigh time \cite{Derakhshani} in all simulated cases, and our numerical computations were validated in Ref. \cite{Carvalho} by replicating some of the experimental results of Seguin et al. \cite{Seguin1}. We present next the results for groups of two or three intruders moving within the granular system.

\section{\label{sec:Res} RESULTS AND DISCUSSION}

\subsection{\label{sec:velocity} Duos moving at constant velocity}

We present in this subsection the results for a pair of aligned intruders moving at constant velocity $V$ = 2.7 mm/s when $\phi$ = 0.76. For different computations, we varied the initial separation $\Delta y$ of intruders and evaluated their drag force $\vec{F}_D$, the mean number of contacts per particle $Z$, the number of non-rattler particles $N$ (particles with at least two contacts), and the anisotropy level $\rho$ ($Z$ and $\rho$ given by the fabric tensor \cite{Bi}, please see the supplementary material for more details). We computed also time averages of these quantities in the entire domain or by considering a moving ROI that followed the intruders along their motion (Fig. \ref{fig:forces_v_cte}(a)). In the case of drag forces, we computed a mean value by averaging the values for both intruders. Those results are summarized in Fig. \ref{fig:forces_v_cte}.

\subsubsection{\label{sec:drag_vel} Drag force on the intruders}

\begin{figure}[h!]
	\begin{center}
		\begin{minipage}{0.49\linewidth}
			\begin{tabular}{c}
				\includegraphics[width=0.80\linewidth]{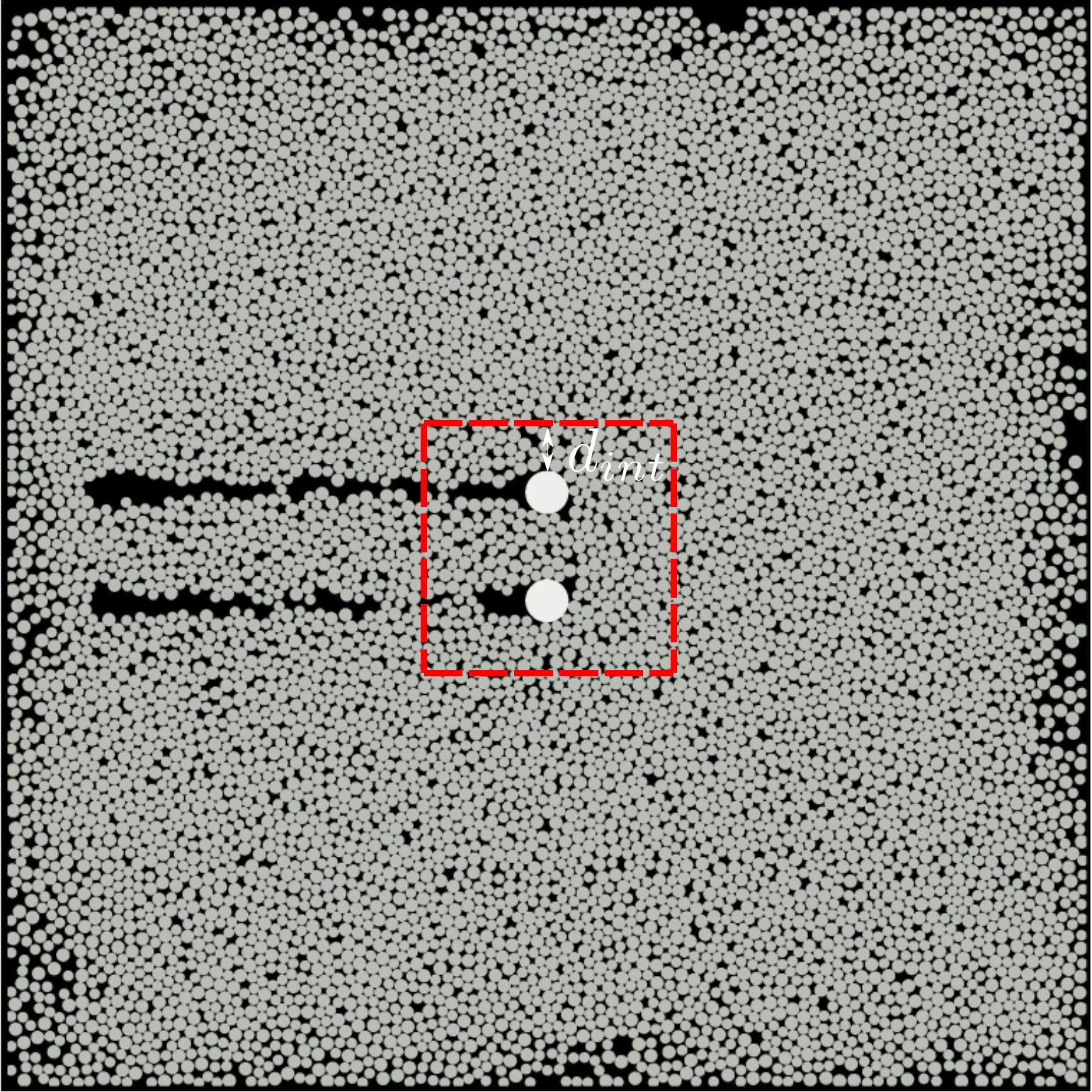}\\
				(a)
			\end{tabular}
		\end{minipage}
		\hfill
		\begin{minipage}{0.49\linewidth}
			\begin{tabular}{c}
				\includegraphics[width=0.80\linewidth]{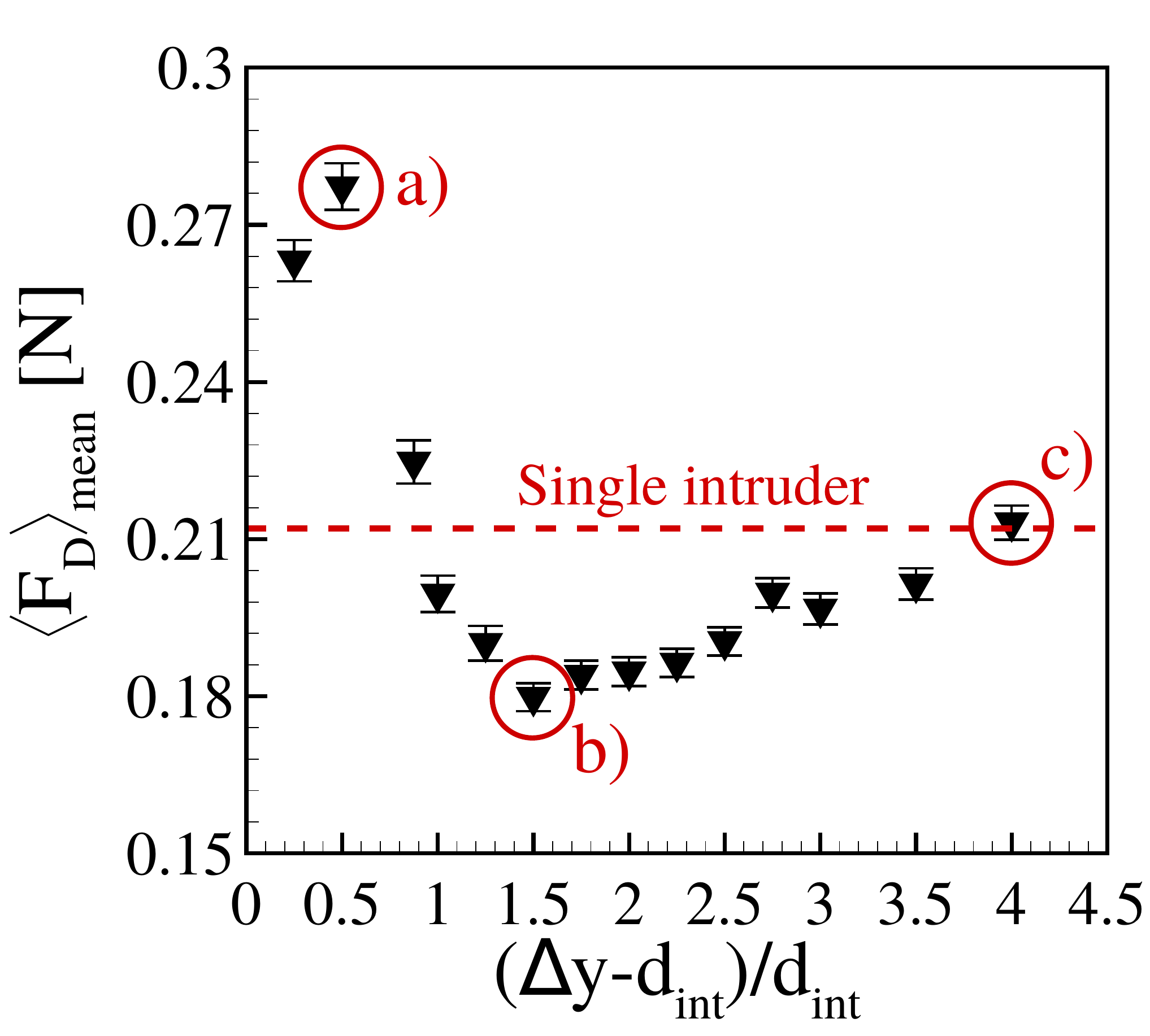}\\
				(b)
			\end{tabular}
		\end{minipage}
		\hfill
		\begin{minipage}{0.49\linewidth}
			\begin{tabular}{c}
				\includegraphics[width=0.80\linewidth]{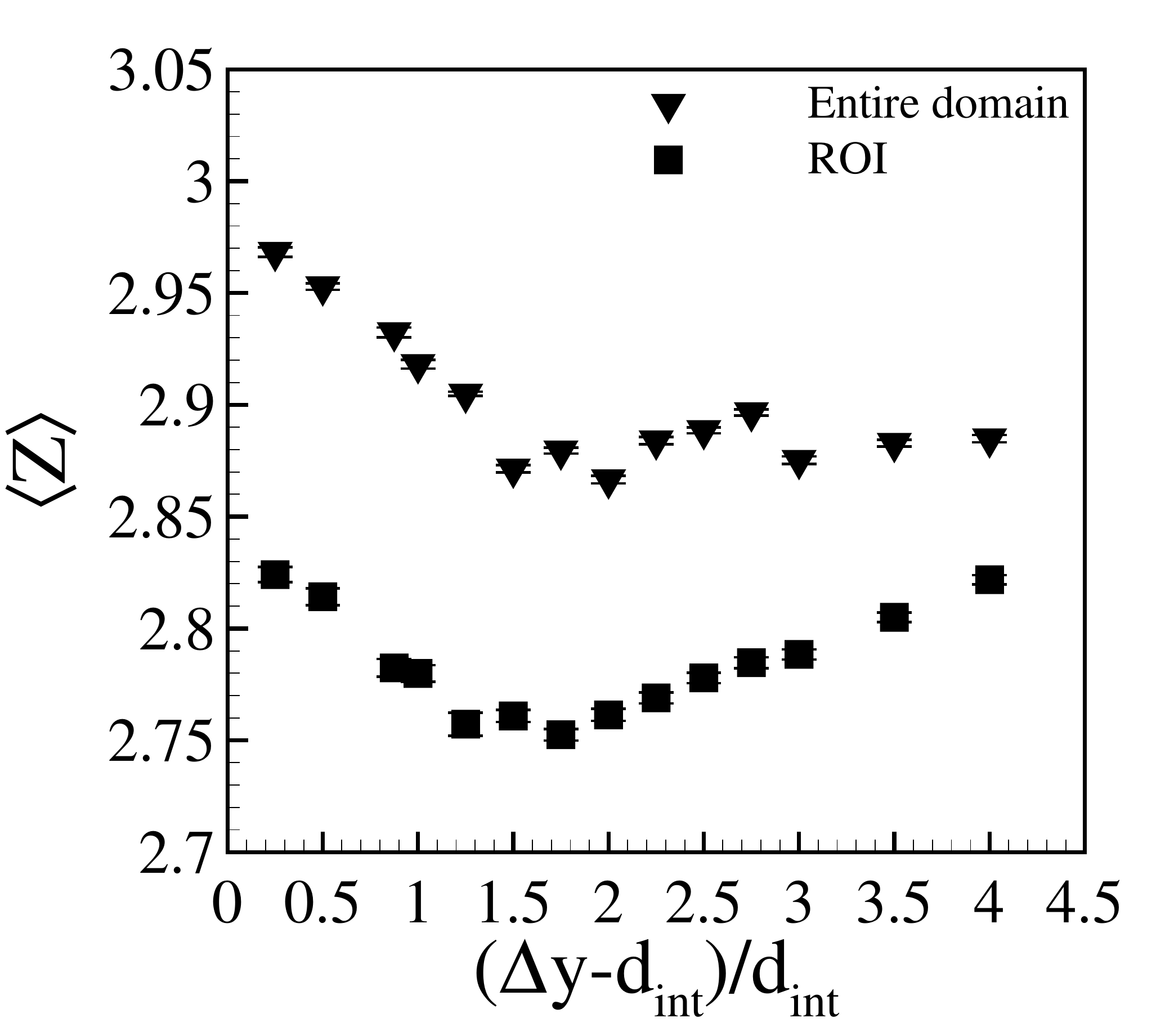}\\
				(c)
			\end{tabular}
		\end{minipage}
		\hfill
		\begin{minipage}{0.49\linewidth}
			\begin{tabular}{c}
				\includegraphics[width=0.80\linewidth]{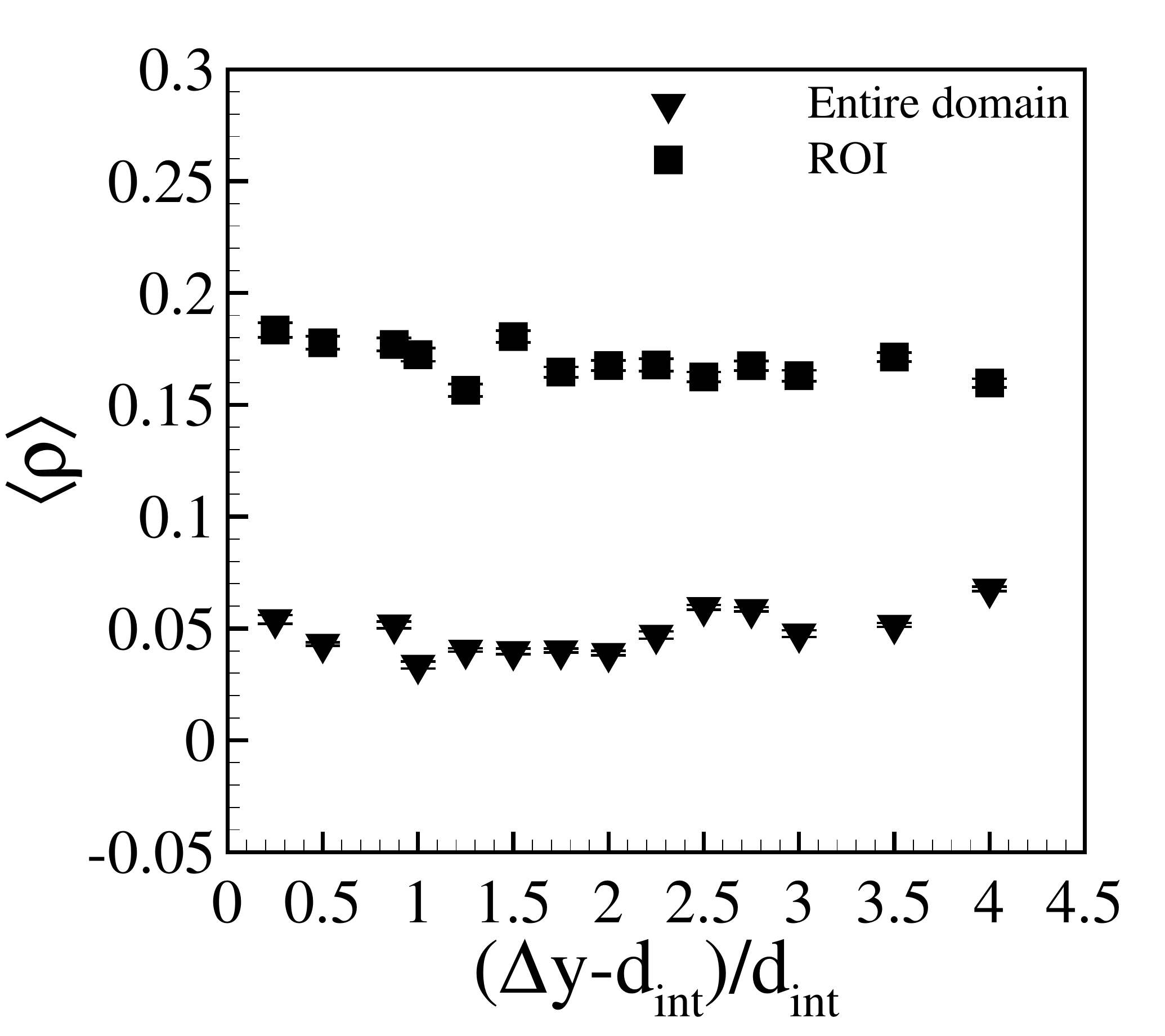}\\
				(d)
			\end{tabular}
		\end{minipage}
		\hfill
	\end{center}
	\caption{(a) Snapshot showing the intruders and grains (top-view image), and a moving ROI around the intruders (delimited by dashed-red lines). (b) Magnitude of the mean resultant force on each intruder $\left< F_D \right>_{mean}$ as a function of their initial separation $\Delta y$ normalized by the intruder diameter $d_{int}$. (c) Mean number of contacts per particle averaged over time $\left< Z \right>$ and (d) time-averaged anisotropy $\left< \rho \right>$ as functions of $(\Delta y - d_{int})/d_{int}$. In figures (b) to (d) triangles correspond to averages computed by considering the entire domain and squares by considering only particles in the moving ROI shown in figure (a), and bars correspond to the standard errors. In figure (b), the dashed-red line corresponds to the time-average drag force found for a single intruder in Ref. \cite{Carvalho}, and the marked points indicate the conditions for which the networks of contact forces are shown in Fig. \ref{fig:contact_chains_duos}. All graphics are for $V$ = 2.7 mm/s and $\phi$ = 0.76.}
	\label{fig:forces_v_cte}
\end{figure}

A snapshot showing a top-view image of grains and intruders for a typical simulation is shown in Fig.\ref{fig:forces_v_cte}(a), where we observe the formation of a cavity (absence of grains) behind (downstream) each intruder in places previously visited by them (movies showing the time evolution of the granular system as the intruders move are available in Figs. \ref{fig:snapshot_trio}, \ref{fig:phase_diagram} and \ref{fig:contact_chains_trios} -- Multimedia view -- and on a public repository \cite{Supplemental2}). The existence of a cavity downstream of a single intruder was measured experimentally by Kolb et al. \cite{Kolb1} and Seguin et al. \cite{Seguin1} and numerically by Carvalho et al. \cite{Carvalho}. The figure also shows the region of interest around the intruders, which moves with them and is used in some computations presented next. The ROI was a square with side length equal to $\Delta y$ $+$ $3d_{int}$.

As in Ref. \cite{Carvalho}, we computed the instantaneous drag force on the intruder $\vec{F}_D$, and then its time-averaged magnitude $\left< F_D \right>$ (examples of the time evolution of $| \vec{F}_D |$ are available in the supplementary material). We afterward computed a mean value for the intruders, $\left< F_D \right>_{mean}$, by adding their time averages and then dividing by two, for each transverse separation $\Delta y$. Figure \ref{fig:forces_v_cte}(b) presents $\left< F_D \right>_{mean}$ as a function of $(\Delta y - d_{int})/d_{int}$, i.e., the separation between the surfaces of intruders normalized by the intruder diameter. We observe a non-monotonic variation, with a decrease in the mean drag force for $(\Delta y - d_{int})/d_{int}$ $<$ 1.5 as the separation between intruders increases, and an increase for $(\Delta y - d_{int})/d_{int}$ $>$ 1.5. Therefore, there is an optimal distance $D_{opt}$ for minimum drag when the separation between the surfaces of intruders is 1.5 times their diameter ($D_{opt}$ $\rightarrow$ $\Delta y$ = 2.5$d_{int}$). Interestingly, in this condition the drag acting on each intruder is approximately 0.18 N, which corresponds to 85\% of the value for a single intruder (Ref. \cite{Carvalho} found 0.21 N for a single intruder under the same velocity and packing fraction). In fact, there is a range 1 $\lessapprox$ $(\Delta y - d_{int})/d_{int}$ $\lessapprox$ 3.5 where $\left< F_D \right>_{mean}$ is smaller than the drag for a single intruder, indicating that some type of cooperative dynamics between the intruders is happening within the granular system. We note that the value of the minimum drag is repeatable, the value 0.18N being obtained in three simulations with different initialization in the same domain (confined by solid walls) and another one with periodic conditions, while a value of approximately 0.19N was obtained for a larger domain ($L_x$ multiplied by 2, and confined by solid walls). Therefore, the results are roughly the same (variations being around 5\% for different domains and initial conditions).

We note that we carried out simulations with different grain sizes (5 and 6 mm), and the results for the time-average drag are similar to previous ones: although the magnitude of the reduction changes with the size of disks, drag reduction occurs for the same range of $(\Delta y - d_{int})/d_{int}$, the minimum drag occurring for $(\Delta y - d_{int})/d_{int}$ $\approx$ 1.5. We thus used $d_{int}$ as a scale for normalization, but further investigation is still necessary. We also carried out simulations with a different velocity of intruders (7.5 mm/s), and obtained similar results: the behavior and regions of drag reduction are the same though the values of drag are different. The results for different sizes of disks and intruder velocity are available in the supplementary material.

\subsubsection{\label{sec:network_vel} Network of contact forces}

\begin{figure}[h!]
	\begin{center}
		\includegraphics[width=0.95\columnwidth]{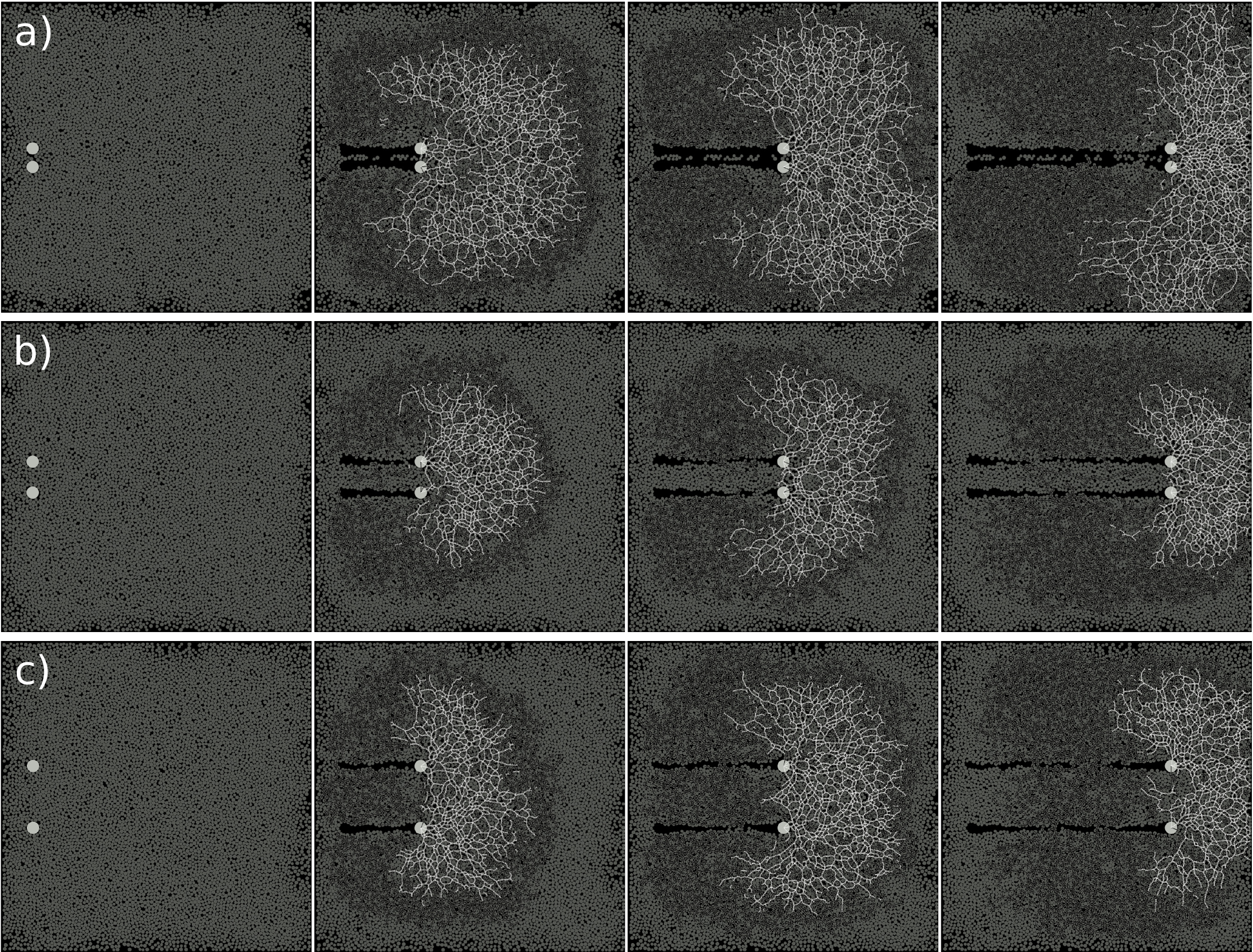}\\
	\end{center}
	\caption{From left to right, snapshots at $t$ = 0, 35.532, 59.220 and 94.752 s of duos moving within grains at $V$ = 2.7 mm/s for (a) $(\Delta y - d_{int})/d_{int}$ = 0.5; (b) $(\Delta y - d_{int})/d_{int}$ = 1.5; and (c) $(\Delta y - d_{int})/d_{int}$ = 4.0. The figures show the load-bearing (clear lines) and dissipative (dark lines) chains for the cases highlighted in Fig. \ref{fig:forces_v_cte}(b), and $\phi$ = 0.76.}
	\label{fig:contact_chains_duos}
\end{figure}

In the case of a cooperative dynamics, we expect that details of the network of contact forces change with the separation between intruders. Following Radjai et al. \cite{Radjai1}, we identified the contact chains and separated them into load-bearing chains, transmitting forces higher than the average value (ensemble average at each considered instant), and dissipative chains, transmitting values lower than that average. The ensemble average was computed at each time step by adding all the contact forces and dividing the result by the number of contacts. Those chains are intermittent, forming and collapsing successively. Figures \ref{fig:contact_chains_duos}(a) to \ref{fig:contact_chains_duos}(c) present snapshots at four different instants of duos moving at constant velocity for, respectively, $(\Delta y - d_{int})/d_{int}$ = 0.5, 1.5, and 4.0, showing also the load-bearing (clear lines) and dissipative (dark lines) chains. The cases $(\Delta y - d_{int})/d_{int}$ = 0.5 and 1.5 correspond, respectively, to the maximum and minimum drags found for $V$ = 2.7 mm/s (Fig. \ref{fig:forces_v_cte}(b)). For $(\Delta y - d_{int})/d_{int}$ = 0.5 (Fig. \ref{fig:contact_chains_duos}(a)), we observe very sporadic load-bearing chains linking the intruders, indicating a low level of (positive) cooperation between the intruders, while load-bearing chains percolate over long distances from the intruders and reach the vertical walls toward the end of their motion. Dissipative chains also reach distances far from the intruders, and by the end of their motion percolate over almost the entire system. For $(\Delta y - d_{int})/d_{int}$ = 1.5 (Fig. \ref{fig:contact_chains_duos}(b)), which corresponds to $D_{opt}$, we notice frequent load-bearing chains linking the intruders, indicating positive cooperation (resulting from one intruder pushing the other via load-bearing chains), and that load-bearing chains remain closer to the intruders, reaching the vertical walls only at the end of motion. Dissipative chains also remain closer to the intruders, reaching the right wall only by the end of their motion (implying less basal drag). For $(\Delta y - d_{int})/d_{int}$ = 4.0, (Fig. \ref{fig:contact_chains_duos}(c)), which corresponds to an intermediate value of $\left< F_D \right>_{mean}$, load-bearing chains link the intruders, but they also extend over distances farther from them. This, together with dissipative chains that percolate also over longer distances and reach the lateral walls before the end of motion, imply larger basal frictions and thus a higher drag force. Therefore, from the balance between chains connecting the intruders and those percolating over long distances, the intruders cooperate the most to move forward when $(\Delta y - d_{int})/d_{int}$ = 1.5 ($\Delta y$ = $D_{opt}$ = 2.5$d_{int}$), pushing each other and being subject to smaller drag.

Because the networks of contact forces are dense and intermittent, it is unfeasible to find small changes directly from their plots, such as those in Fig. \ref{fig:contact_chains_duos}. We thus investigated how the mean number of contacts per particle averaged over time $\left< Z \right>$ and the time-averaged anisotropy $\left< \rho \right>$ change with $(\Delta y - d_{int})/d_{int}$, which we show in Figs. \ref{fig:forces_v_cte}(c) and \ref{fig:forces_v_cte}(d), respectively, computed for both the entire domain and the ROI.

By considering the entire domain, we observe that the profile of $\left< Z \right>$ follows roughly that of $\left< F_D \right>_{mean}$, with $\left< Z \right>$ $\approx$ 2.9 in the region 1 $\leq$ $(\Delta y - d_{int})/d_{int}$ $\leq$ 4, which is approximately the value for the single intruder (2.903) \cite{Carvalho}. The anisotropy $\left< \rho \right>$ considering the entire domain is roughly constant, with values of the order of 0.05, also close to that for the single intruder (0.041) \cite{Carvalho}. Therefore, the granular system as a whole does not seem to play a large role on drag reduction, except when the intruders are too close from each other ($(\Delta y - d_{int})/d_{int}$ $<$ 1), being in that case almost one single and large intruder. By considering now only the ROI, the general behaviors of both $\left< Z \right>$ and $\left< \rho \right>$ with $(\Delta y - d_{int})/d_{int}$ remain as for the entire region, but the magnitudes are different: $\left< Z \right>$ varies around 2.8 (smaller than that for the single intruder) and $\left< \rho \right>$ remains constant at approximately 0.1 (larger than that for the single intruder). The network of contact forces is thus different in the neighborhood of the intruders, with less contacts between grains and higher anisotropy, indicating preferential directions for percolating loads. These preferential directions are connected with the motion of disks shown in Subsection \ref{sec:motion_disks}.

Our results show that: (i) there exists a cooperative dynamics between the intruders; and (ii) in cases of constant velocity, there is an optimal separation between intruders for not only reaching minimum drag, but also drag reduction (with respect to single intruders). The latter can be proven useful for designing devices stirring the ground or other granular surfaces.

Additional graphics for the number of contacts per particle $Z$, number of non-rattler particles $N$ and anisotropy $\rho$ are available in the supplementary material.

\subsection{\label{sec:force} Constant thrusting force}

We investigate now the behavior of the entire system when a constant external force (thrust) of 0.8 N in the longitudinal direction is imposed on each intruder of either duos or trios ($\phi$ = 0.76, unless where otherwise mentioned). In these cases, because the drag force oscillates along the motion and the intruders are free to move in the transverse direction, the cooperative behavior implies intruder velocities that vary along time and depend on the initial configuration, giving rise to different types of migration.

\subsubsection{\label{sec:patterns} Patterns: collaborative motion}

\begin{figure}[h!]
	\begin{center}
		\includegraphics[width=0.95\columnwidth]{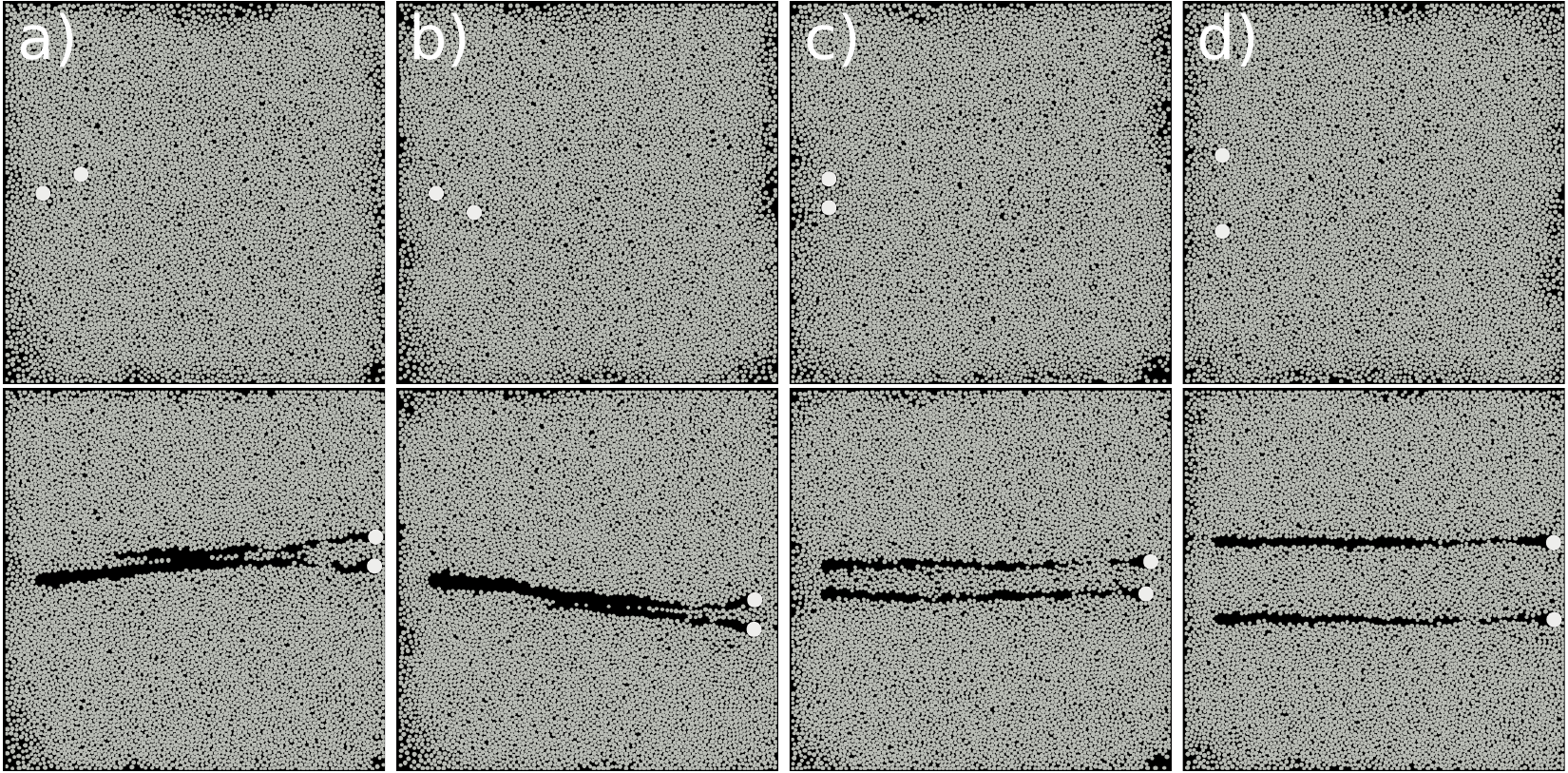}\\ 
	\end{center}
	\caption{Snapshots showing the intruders and grains (top-view images) for a pair of intruders: (a) and (b) off-centered (type III); and (c) and (d) aligned (type IV) in transverse direction. The initial condition is on the top and the final configuration on the bottom of each subfigure, and $\phi$ = 0.76.}
	\label{fig:snapshot_duo}
\end{figure}

By varying the initial configuration of duos and trios, both in terms of orientation and separations, we obtained different migration characteristics that can be classified in patterns. Beginning by duos, Fig. \ref{fig:snapshot_duo} shows the investigated cases, which consisted of intruders aligned (type IV, Figs. \ref{fig:snapshot_duo}(c) and \ref{fig:snapshot_duo}(d)), and off-centered (type III, Figs. \ref{fig:snapshot_duo}(a) and \ref{fig:snapshot_duo}(b), which are symmetrical) in the transverse direction, with the initial condition on the top and the final on the bottom of each subfigure. We observe basically three patterns: (i) when the intruders are off-centered (type III), the cavity generated by the one that is in front (upstream) affects the intruder that is behind (downstream), which then moves faster in longitudinal direction with a component in the transverse direction toward the upstream intruder. Both intruders migrate with a small transverse component, and by the end  of their motion are aligned in the transverse direction, separated by a characteristic distance $D_{att}$ (Figs. \ref{fig:snapshot_duo}(a) and \ref{fig:snapshot_duo}(b)). (ii) For aligned intruders (type IV) with $\Delta y$ within a certain range, they approach or retreat until reaching $\Delta y_{final}$ = $D_{att}$, and move afterward in aligned configuration, keeping $\Delta y$ = $D_{att}$ (Fig. \ref{fig:snapshot_duo}(c)). (iii) For aligned intruders (type IV) with $\Delta y$ above a given threshold, the intruders move in aligned configurations maintaining $\Delta y$ approximately constant (Fig. \ref{fig:snapshot_duo}(d)). The temporal evolution of $\Delta y$ for the cases of Fig. \ref{fig:snapshot_duo} are available in the supplementary material. We note that the evolution toward $D_{att}$ depends on the presence of solid boundaries: we computed one case with periodic boundaries and observed that, although there is still a collaborative motion, the grains did not evolve to $D_{att}$ (at least in the simulated domain). However, we investigate here the confined case, and, therefore, we present next only the cases with solid boundaries.

\begin{figure}[h!]
	\begin{center}
		\includegraphics[width=0.95\columnwidth]{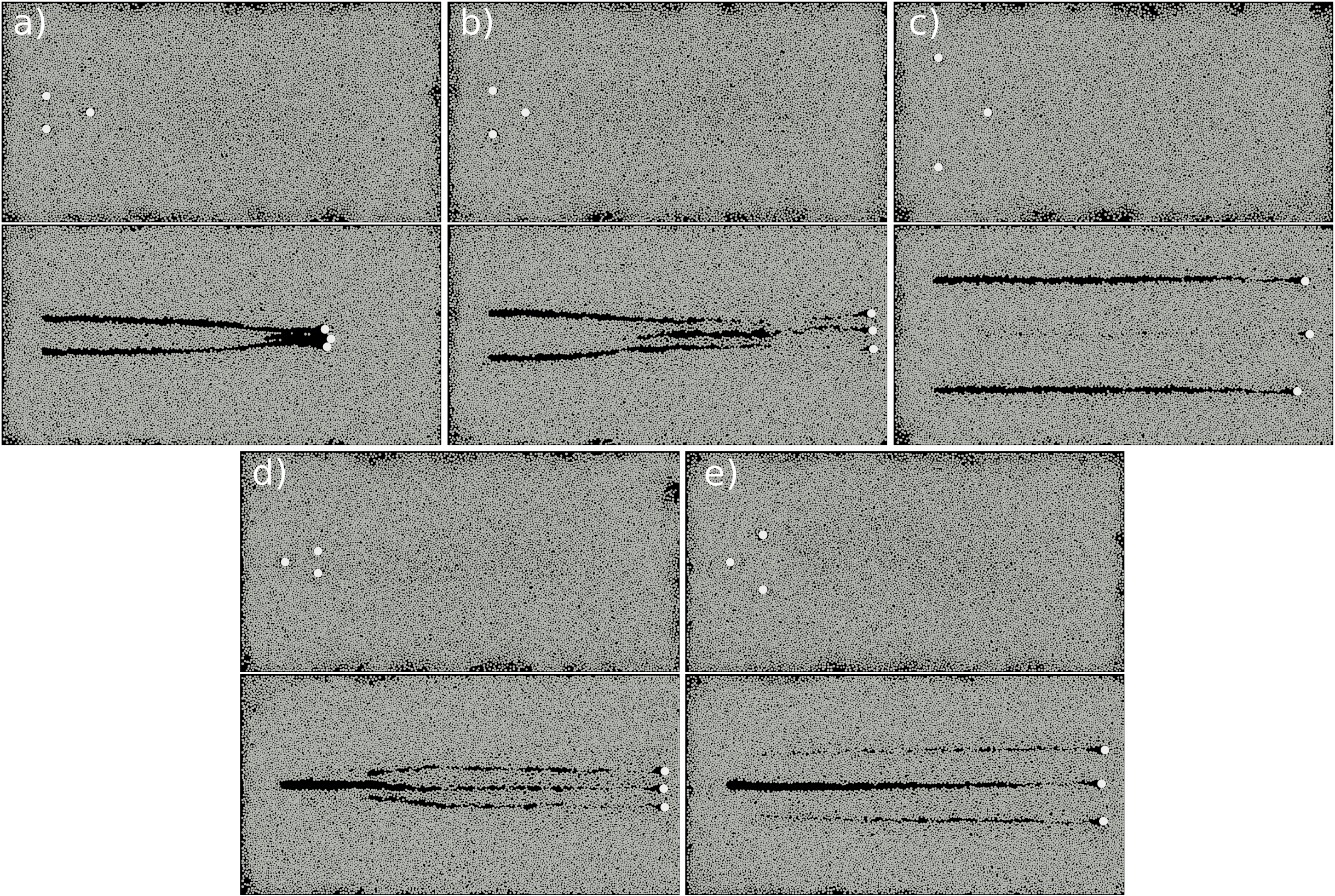}\\
	\end{center}
	\caption{Snapshots showing the intruders and grains (top-view images), for (a), (b) and (c) one intruder in front of two intruders initially aligned in the transverse direction (type I) and (d) and (e) one intruder behind two intruders initially aligned in the transverse direction (type II). The initial condition is on the top and the final configuration on the bottom of each subfigure, and $\phi$ = 0.76. Multimedia view}
	\label{fig:snapshot_trio}
\end{figure}

For the trios, Fig. \ref{fig:snapshot_trio} (Multimedia view) shows the cases that we investigated, namely one intruder in front of two intruders initially aligned in the transverse direction (type I, Figs. \ref{fig:snapshot_trio}(a) to \ref{fig:snapshot_trio}(c)) and one intruder behind two intruders initially aligned in the transverse direction (type II, Figs. \ref{fig:snapshot_trio}(d) and \ref{fig:snapshot_trio}(e)). We observe basically three patterns: (i) for type I with small separations (small values of $\Delta y$), the downstream intruders are exposed to the cavity of the upstream one, and thus move faster in the longitudinal direction and toward the upstream intruder in the transverse direction, forming a clump at the end of their motion (Fig. \ref{fig:snapshot_trio}(a)). At that time, the drag force is highly increased so that the intruders stop, with indications of jamming (described in Subsection \ref{sec:network}). (ii) Within a certain range of $\Delta y$ in types I and II (for type II it is $\Delta y$ lower than a given threshold), the downstream intruder(s) move(s) faster in the longitudinal direction and end(s) finally aligned with the upstream one(s), while in the transverse direction they move until reaching $\Delta y_{final}$ = $D_{att}$ (Figs. \ref{fig:snapshot_trio}(b) and \ref{fig:snapshot_trio}(d)). (iii)  For $\Delta y$ above a certain value (large separations) in types I and II, the downstream intruder(s) move(s) faster in the longitudinal direction and end(s) finally aligned with the upstream one without changing the transverse separation ($\Delta y$ remains approximately constant, Figs. \ref{fig:snapshot_trio}(c) and \ref{fig:snapshot_trio}(e)).

\begin{figure}[h!]
	\begin{center}
		\includegraphics[width=0.4\columnwidth]{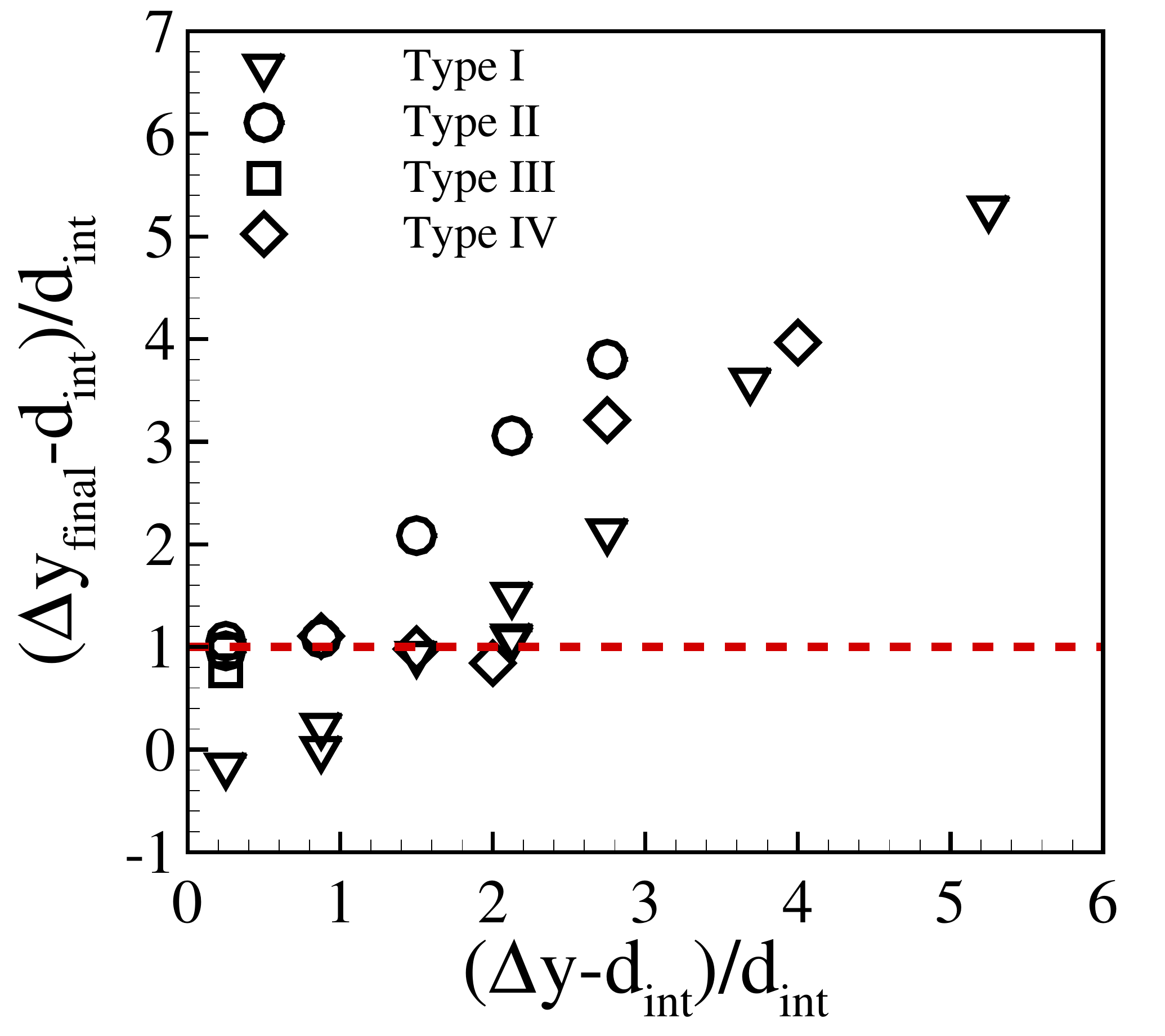}\\
	\end{center}
	\caption{Chart for collaborative patterns: final separations $\Delta y_{final}$ as functions of initial separations $\Delta y$. The symbols are listed in the key and the dashed-red line corresponds to $(\Delta y_{final} - d_{int})/d_{int}$ = 1. Multimedia view}
	\label{fig:phase_diagram}
\end{figure}

The different patterns are summarized in Fig. \ref{fig:phase_diagram}, which shows the final separations $\Delta y_{final}$ as functions of the initial ones $\Delta y$, normalized by the diameter of intruders $d_{int}$. In this figure, the dashed-red line corresponds to $(\Delta y_{final} - d_{int})/d_{int}$ = 1, and the symbols are listed in the figure key. We observe in Fig. \ref{fig:phase_diagram} the behaviors described in previous paragraphs, but we can now find the respective ranges of initial separations and the value of $D_{att}$. For type I, within 1.5 $\leq$ $(\Delta y - d_{int})/d_{int}$ $\leq$ 2.2 the final distances reach $(\Delta y_{final} - d_{int})/d_{int}$ $\approx$ 1 (i.e., $D_{att}$ = 2$d_{int}$). For smaller values of $\Delta y$, the intruders form a clump ($\Delta y_{final}$ $<$ $\Delta y$) and for higher values they keep their separation ($\Delta y_{final}$ $\approx$ $\Delta y$). For type II with $(\Delta y - d_{int})/d_{int}$ $\leq$ 1 and types III and IV with $(\Delta y - d_{int})/d_{int}$ $\leq$ 2, the intruders reach $D_{att}$ = 2$d_{int}$, while for $(\Delta y - d_{int})/d_{int}$ $>$ 1 or 2 (for types II or III and IV, respectively) the transverse separations remain constant ($\Delta y_{final}$ $\approx$ $\Delta y$). Therefore, within certain ranges of initial separations a fixed $\Delta y_{final}$ is reached (an attractor-like behavior), corresponding to $D_{att}/d_{int}$ = 2 (surface-surface separations equal to $d_{int}$). We note that one of the runs for type I within 1.5 $\leq$ $(\Delta y - d_{int})/d_{int}$ $\leq$ 2.2 does not fall exactly on $(\Delta y_{final} - d_{int})/d_{int}$ $\approx$ 1, but lies close to it. We consider this deviation rather small, resuting from the presence of one additional disk between two of the intruders at the final stage.

\subsubsection{\label{sec:motion_disks} Motion of disks}

\begin{figure}[h!]
	\begin{center}
		\includegraphics[width=0.95\columnwidth]{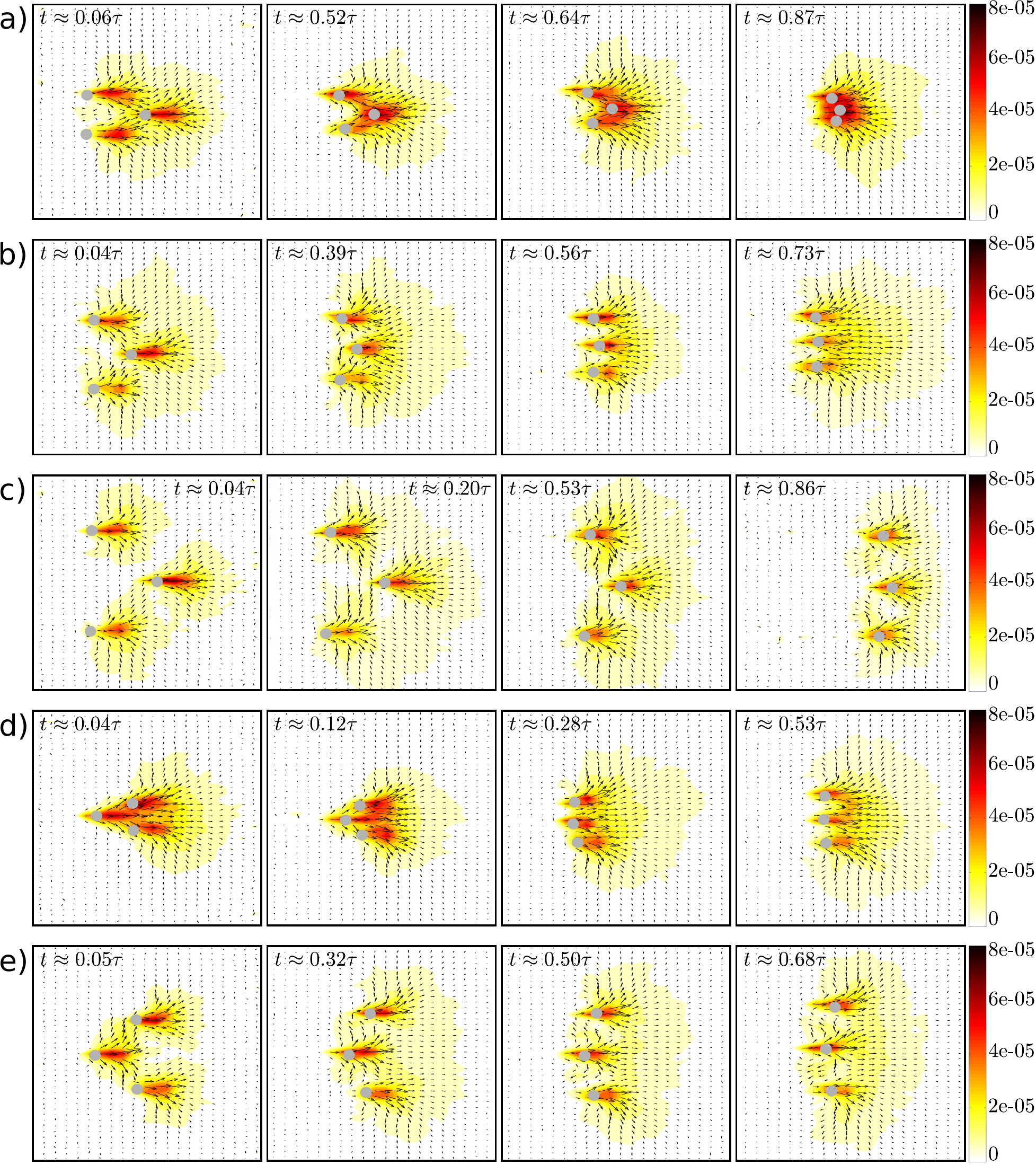}\\
	\end{center}
	\caption{Snapshots showing the velocity fields of disks as the intruders move, for (a), (b) and (c) one intruder in front of two intruders initially aligned in the transverse direction (type I) and (d) and (e) one intruder behind two intruders initially aligned in the transverse direction (type II). $\tau$ is the total time of the intrudes' motion, being 2064, 2796, 2928, 2964 and 2652 s for figures (a) to (e), respectively. The cases are in the same order of Fig. \ref{fig:snapshot_trio} ($\phi$ = 0.76), and the scale of the colorbar is in m/s.}
	\label{fig:displacements_disks}
\end{figure}

The collaborative behavior can be examined in terms of motion of disks as the intruders are thrust within them. Figure \ref{fig:displacements_disks} shows snapshots of the velocity fields of disks for one intruder in front of two intruders initially aligned in the transverse direction (type I, Figs. \ref{fig:displacements_disks}(a) to \ref{fig:displacements_disks}(c)) and one intruder behind two intruders initially aligned in the transverse direction (type II, Figs.  \ref{fig:displacements_disks}(d) and \ref{fig:displacements_disks}(e)). In Fig. \ref{fig:displacements_disks}, $\phi$ = 0.76 (corresponds to the same cases presented in Fig. \ref{fig:snapshot_trio}) and $\tau$ is the total time of the intruders' motion. With the exception of Fig. \ref{fig:displacements_disks}(a) (clump), we observe that even if disks just in front of the intruder have higher velocities, the motion reaches regions far from the intruders, extending toward the walls by the end of the intruders' motion. Throughout the motion, we can also observe grains recirculating from the intruders' front toward their rear, and also migrating from the compacted regions toward the cavities. The migration toward cavities is particularly noticeable in Fig. \ref{fig:displacements_disks}(c), where vectors indicate migration to the central cavity and, indeed, Fig. \ref{fig:snapshot_trio}(c) shows that this cavity is almost suppressed, and in Fig. \ref{fig:displacements_disks}(e), where vectors indicate migration to the lateral cavities, also shown in Fig. \ref{fig:snapshot_trio}(e) (where those cavities are partially suppressed).

In the specific case of Fig. \ref{fig:displacements_disks}(a) (clump), during roughly the first half of motion ($t$ $\approx$ 0.06$\tau$ to $t$ $\approx$ 0.64$\tau$) part of disks in front the downstream intruders move toward the upstream intruder and suppress its cavity, as can be seen in Fig. \ref{fig:snapshot_trio}(a). During this time, disks recirculate around the downstream intruders, migrating from compacted regions toward their cavities. Close to the end of motion ($t$ $\approx$ 0.87$\tau$), we observe that the intruders are close to each other and the motion of disks is concentrated just in front of them, not reaching regions so far from the intruders as in the other four cases, indicating a compacted region in front of them. At the same time, a large cavity forms behind the intruders (shown in Fig. \ref{fig:snapshot_trio}(a)), the degree of velocities being small in the recirculation region. All that indicates that the intruders are about to be blocked.

In common, those cases show that part of grains in front of downstream intruders are pushed toward the cavity in front of them, generated by the upstream intruder. This facilitates the motion of downstream intruders, which move faster and reach eventually the upstream ones, whether to be transversely aligned or to form a clump. This picture is in agreement with experiments conducted with a single intruder moving in a system of disks with the same size as in our simulations. For instance, Kolb et al. \cite{Kolb1} and Seguin et al. \cite{Seguin1} showed that a compacted region forms in front of the intruder, with disks reaching the packing fraction for jamming, while a decompressed region, usually a cavity, forms behind the intruder, so that disks recirculate from the compacted front toward the cavity. The recirculation is intermittent, making the drag force to fluctuate around a mean value (as we showed in Carvalho et al. \cite{Carvalho} and in Subsection \ref{sec:drag_force}).

\subsubsection{\label{sec:drag_force} Drag force on the intruders}

\begin{table}[h!]
	\centering
	\caption{Mean drag on each intruder for different types and separations: configuration type, initial separation in the longitudinal direction $\Delta x$, initial separation in the transverse direction $\Delta y$, and average forces on intruders 1, 2 and 3, $\left< F_D \right>_1$, $\left< F_D \right>_2$ and $\left< F_D \right>_3$, respectively. Values are normalized by $d_{int}$ and $F_T$.}
	\begin{tabular}{|c|c|c|c|c|c|}
		\hline
		Type & $\Delta x / d_{int}$ & $\Delta y / d_{int}$ & $\left< F_D \right>_1 / F_T$ & $\left< F_D \right>_2 / F_T$ & $\left< F_D \right>_3 / F_T$ \\ \hline
		I    & 5.00   	& 1.88  & 0.54   & 0.46   & 0.46  \\ \hline
		I    & 3.75 & 1.25  & 0.53   & 0.43   & 0.44   \\ \hline
		I    & 3.75 & 1.88  & 0.54   & 0.44   & 0.45   \\ \hline
		I    & 5.00 & 3.13  & 0.55   & 0.49   & 0.49   \\ \hline
		I    & 3.75 & 2.50  & 0.61   & 0.58   & 0.57   \\ \hline
		I    & 3.75 & 3.13  & 0.51   & 0.47   & 0.47   \\ \hline
		I    & 5.63 & 3.13  & 0.54   & 0.47   & 0.47   \\ \hline
		I    & 5.63 & 4.69  & 0.55   & 0.49   & 0.49   \\ \hline
		I    & 5.63 & 6.25  & 0.56   & 0.50   & 0.51   \\ \hline
		I    & 3.75 & 3.75  & 0.52   & 0.47   & 0.47   \\ \hline
		II   & 3.75 & 1.25  & 0.49   & 0.53   & 0.53   \\ \hline
		II   & 3.75 & 1.88  & 0.47   & 0.51   & 0.51   \\ \hline
		II   & 6.25 & 1.25  & 0.48   & 0.56   & 0.56   \\ \hline
		II   & 3.75 & 3.13  & 0.43   & 0.48   & 0.48   \\ \hline
		II   & 6.25 & 2.50  & 0.40   & 0.48   & 0.48   \\ \hline
		II   & 6.25 & 3.75  & 0.44   & 0.50   & 0.51   \\ \hline
		III  & 2.50 & 1.25  & 0.57   & 0.61   & -      \\ \hline
		III  & 2.50 & 1.25  & 0.42   & 0.54   & -      \\ \hline
		III  & 2.50 & 1.25  & 0.49   & 0.55   & -      \\ \hline
		IV   & 0    & 1.88  & 0.53   & 0.52   & -      \\ \hline
		IV   & 0    & 2.50  & 0.49   & 0.49   & -      \\ \hline
		IV   & 0    & 3.00  & 0.53   & 0.52   & -      \\ \hline
		IV   & 0    & 3.75  & 0.49   & 0.49   & -      \\ \hline
		IV   & 0    & 5.00  & 0.54   & 0.55   & -      \\ \hline
	\end{tabular}
	\label{tab:mean_force}
\end{table} 

We measured the magnitude of the instantaneous drag force on each intruder by computing $F_{D}$ = $\sqrt{(F_{x}-F_{T})^{2} + F_{y}^2}$, where $F_{T}$ = 0.8 N is the magnitude of the thrusting force imposed on each intruder, and $F_x$ and $F_y$ are the longitudinal and transverse components of the resultant force (so that the basal friction is included in $F_D$). We afterward time-averaged $F_{D}$ for each intruder, obtaining $\left< F_D \right>$ for different initial separations. The results are summarized in Tab. \ref{tab:mean_force}, where $\left< F_D \right>_1$, $\left< F_D \right>_2$ and $\left< F_D \right>_3$ correspond to time-averaged drag forces on intruders labeled 1, 2 and 3, respectively, as shown in Fig. \ref{fig:setup}. Distances in Tab. \ref{tab:mean_force} are normalized by the diameter of intruders, $d_{int}$, and forces by the thrusting force on each intruder, $F_{T}$. From Tab. \ref{tab:mean_force}, we notice basically that the drag forces on the upstream intruders are larger than those on the downstream ones, being roughly equal for intruders aligned in the transverse direction (graphics of the mean drag forces as functions of initial separations are available in the supplementary material). This corroborates the description given in the previous subsection. We note that the initial separation in the longitudinal direction ($\Delta x$) only affects the time for reaching the final configuration (please see the supplementary material for a graphic of the time to reach the final configuration as a function of the initial separation between intruders).

\subsubsection{\label{sec:network} Network of contact forces}

\begin{figure}[h!]
	\begin{center}
		\includegraphics[width=0.95\columnwidth]{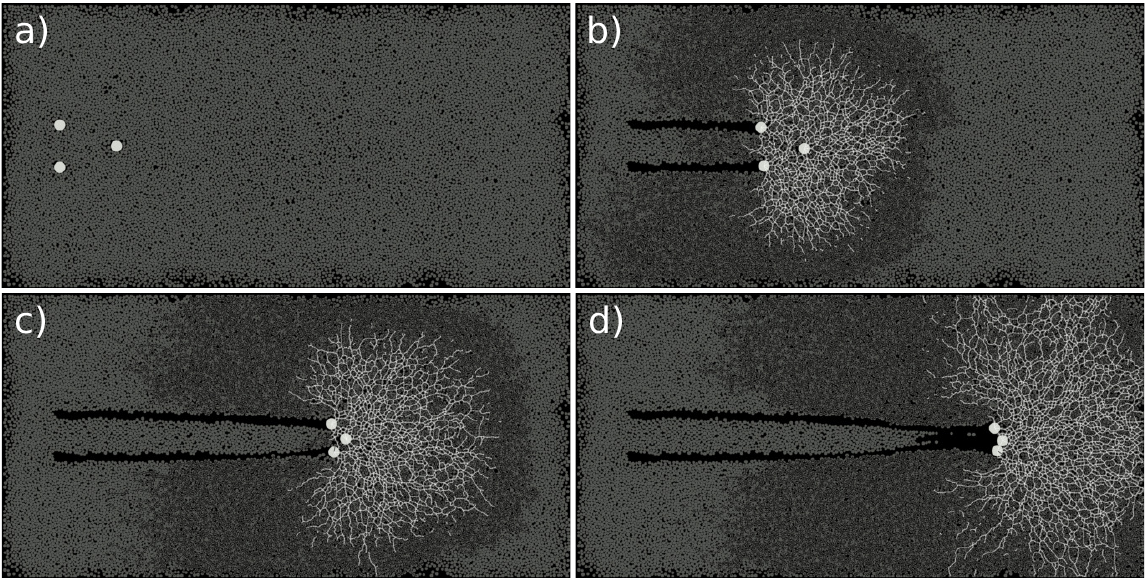}\\
	\end{center}
	\caption{From left to right, snapshots at (a) $t$ = 0 s, (b) $t$ = 710.64 s, (c) $t$ = 1492.80 s and (d) $t$ = 2074.56 s for trios organized in type I being pushed at 0.8 N (each intruder, same of Fig. \ref{fig:snapshot_trio}(a)). The figures show the load-bearing (clear lines) and dissipative (dark lines) chains, and $\phi$ = 0.76. Multimedia view}
	\label{fig:contact_chains_trios}
\end{figure}

We examine now how the network of contact forces is related with drag reduction, increase, and even jamming in certain cases. For that, we proceeded as in Subsection \ref{sec:network_vel} and computed the load-bearing and dissipative chains. Figure \ref{fig:contact_chains_trios} (Multimedia view) shows the load-bearing and dissipative chains superposed with the particle positions at four different instants for the same case of Fig. \ref{fig:snapshot_trio}(a) (the three intruders form a clump which jams). Although load-bearing chains exist in the region between the intruders, we notice that they percolate over longer distances as the intruders come closer to each other. By the end of the intruders' motion (Fig. \ref{fig:contact_chains_trios}(d)), load-bearing chains are dense and reach three of the vertical walls, blocking the motion of the intruders (and indicating a possible jamming state.).

Graphics of the evolution of $Z$ and $\rho$ are available in the supplementary material, and a movie showing the motion of grains and force networks within the granular system is available in Figure \ref{fig:contact_chains_trios} (Multimedia view).

\subsubsection{\label{sec:packing} Variations with the packing fraction}

\begin{figure}[h!]
	\begin{center}
		\includegraphics[width=0.95\columnwidth]{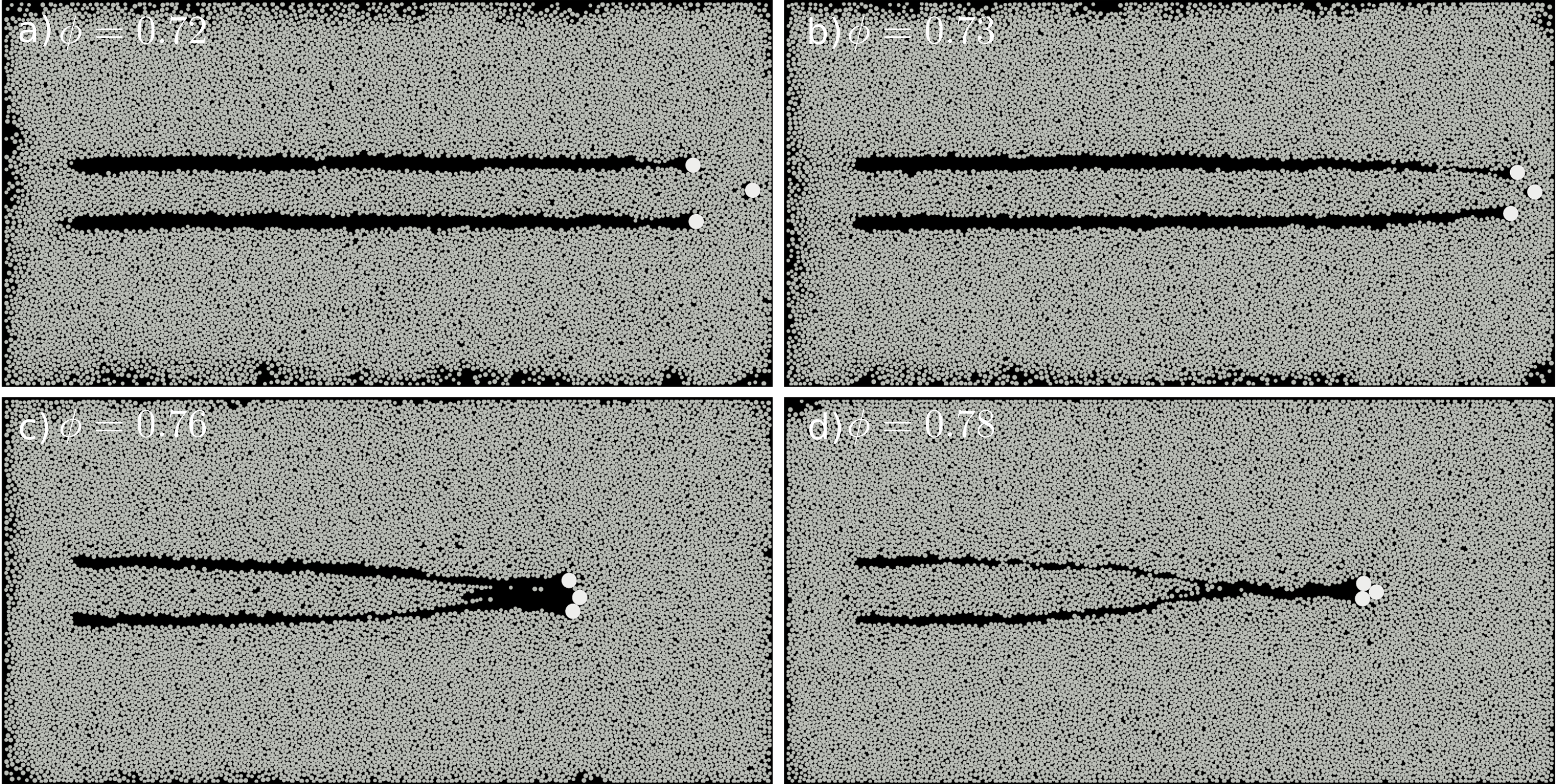}\\
	\end{center}
	\caption{Snapshots of final positions for simulations of type I with (a) $\phi$ = 0.72, (b) $\phi$ = 0.73, (c) $\phi$ =0.76, and (d) $\phi$ =0.78 (same initial configuration of intruders as in Fig. \ref{fig:snapshot_trio}(a)).}
	\label{fig:snapshot_duo_phivar}
\end{figure}

We investigate in this subsection if some of the patterns shown previously for $\phi$ = 0.76 change with the packing fraction. For example, Figs. \ref{fig:snapshot_duo_phivar}(a) to \ref{fig:snapshot_duo_phivar}(d) show snapshots of final positions for simulations of type I with $\phi$ = 0.72, 0.73, 0.76 and 0.78, respectively, for the same initial separations (same initial configuration of intruders as in Fig. \ref{fig:snapshot_trio}(a), $(\Delta y - d_{int})/d_{int}$ = 0.875). We observe that the increase in packing fraction boosts the cooperative dynamics between intruders: for the lowest value ($\phi$ = 0.72), the intruders move as single objects, with virtually no cooperation, while from $\phi$ = 0.72 to $\phi$ = 0.78 they tend to form the clumped structure in a time scale that decreases as $\phi$ increases. Although we  show here that $\phi$ is an important parameter in determining the different patterns observed, we do not inquire into its effects, which needs to be investigated further.

\section{\label{sec:Conclu} CONCLUSIONS}

In this paper, we investigated how a group of intruders interact with each other while moving horizontally in a two-dimensional granular system. Our results show that: (i) there exists a cooperative dynamics between the intruders; (ii) this cooperative dynamics is the result of compaction and expansion of the granular medium in front and behind, respectively, each intruder, with load-bearing chains connecting the intruders and cavities being formed in front of the downstream intruders; (iii) for the cases presenting more drag, load-bearing chains percolate over longer distances, reaching in some cases the vertical walls; (iv) in cases of constant velocity, there is an optimal separation between intruders for not only reaching minimum drag, but also drag reduction (with respect to single intruders). This can be proven useful for designing devices stirring the ground or other granular surfaces; (v) for constant thrust, different patterns appear depending on the initial configurations and distances between intruders; (vi) in addition to initial separations, the packing fraction also influences the observed patterns; (vii) for some initial arrangements, the same spatial configuration is eventually reached, showing an attractor-like behavior. While a cooperative dynamics was shown in the case of intruders falling within light grains by Refs. \cite{Pacheco, Solano, Dhiman, Kawabata, Pravin, Espinosa}, and the compaction and expansion of the granular medium in front and behind a single intruder by Refs. \cite{Kolb1,Seguin1}, all the remaining findings are new. However, despite the progress made, some questions remain to be investigated further, such as the variation of the magnitude of the drag reduction with the sizes of disks and/or intruders and with the intruders' velocity, the influence of the shapes of disks and intruders (elliptical, angular, etc.) on the dynamics of the entire system, and the relation between the packing fraction and the optimal distance for minimum drag. On the whole, our results bring new insights into the cooperative dynamics of intruders moving amid grains.

\section*{AUTHOR DECLARATIONS}
\noindent \textbf{Conflict of Interest}

The authors have no conflicts to disclose

\section*{SUPPLEMENTARY MATERIAL}
See supplementary material for additional graphics for the remaining numerical data.

\section*{DATA AVAILABILITY}
The data that support the findings of this study are openly available in Mendeley Data at https://data.mendeley.com/datasets/39wgn3jtxb.

\begin{acknowledgments}
The authors are grateful to FAPESP (Grant Nos. 2018/14981-7 and 2020/04151-7) for the financial support provided. The authors would like to thank N. C. Lima for the help with the initial setup and F. D. C\'u\~nez for fruitful discussions.
\end{acknowledgments}

\bibliography{references}

\begin{thebibliography}{43}%
\makeatletter
\providecommand \@ifxundefined [1]{%
 \@ifx{#1\undefined}
}%
\providecommand \@ifnum [1]{%
 \ifnum #1\expandafter \@firstoftwo
 \else \expandafter \@secondoftwo
 \fi
}%
\providecommand \@ifx [1]{%
 \ifx #1\expandafter \@firstoftwo
 \else \expandafter \@secondoftwo
 \fi
}%
\providecommand \natexlab [1]{#1}%
\providecommand \enquote  [1]{``#1''}%
\providecommand \bibnamefont  [1]{#1}%
\providecommand \bibfnamefont [1]{#1}%
\providecommand \citenamefont [1]{#1}%
\providecommand \href@noop [0]{\@secondoftwo}%
\providecommand \href [0]{\begingroup \@sanitize@url \@href}%
\providecommand \@href[1]{\@@startlink{#1}\@@href}%
\providecommand \@@href[1]{\endgroup#1\@@endlink}%
\providecommand \@sanitize@url [0]{\catcode `\\12\catcode `\$12\catcode
  `\&12\catcode `\#12\catcode `\^12\catcode `\_12\catcode `\%12\relax}%
\providecommand \@@startlink[1]{}%
\providecommand \@@endlink[0]{}%
\providecommand \url  [0]{\begingroup\@sanitize@url \@url }%
\providecommand \@url [1]{\endgroup\@href {#1}{\urlprefix }}%
\providecommand \urlprefix  [0]{URL }%
\providecommand \Eprint [0]{\href }%
\providecommand \doibase [0]{https://doi.org/}%
\providecommand \selectlanguage [0]{\@gobble}%
\providecommand \bibinfo  [0]{\@secondoftwo}%
\providecommand \bibfield  [0]{\@secondoftwo}%
\providecommand \translation [1]{[#1]}%
\providecommand \BibitemOpen [0]{}%
\providecommand \bibitemStop [0]{}%
\providecommand \bibitemNoStop [0]{.\EOS\space}%
\providecommand \EOS [0]{\spacefactor3000\relax}%
\providecommand \BibitemShut  [1]{\csname bibitem#1\endcsname}%
\let\auto@bib@innerbib\@empty
\bibitem [{\citenamefont {Kolb}\ \emph {et~al.}(2013)\citenamefont {Kolb},
  \citenamefont {Cixous}, \citenamefont {Gaudouen},\ and\ \citenamefont
  {Darnige}}]{Kolb1}%
  \BibitemOpen
  \bibfield  {author} {\bibinfo {author} {\bibfnamefont {E.}~\bibnamefont
  {Kolb}}, \bibinfo {author} {\bibfnamefont {P.}~\bibnamefont {Cixous}},
  \bibinfo {author} {\bibfnamefont {N.}~\bibnamefont {Gaudouen}},\ and\
  \bibinfo {author} {\bibfnamefont {T.}~\bibnamefont {Darnige}},\ }\bibfield
  {title} {\enquote {\bibinfo {title} {Rigid intruder inside a two-dimensional
  dense granular flow: Drag force and cavity formation},}\ }\href
  {https://doi.org/10.1103/PhysRevE.87.032207} {\bibfield  {journal} {\bibinfo
  {journal} {Phys. Rev. E}\ }\textbf {\bibinfo {volume} {87}},\ \bibinfo
  {pages} {032207} (\bibinfo {year} {2013})}\BibitemShut {NoStop}%
\bibitem [{\citenamefont {Tordesillas}, \citenamefont {Hilton},\ and\
  \citenamefont {Tobin}(2014)}]{Tordesillas}%
  \BibitemOpen
  \bibfield  {author} {\bibinfo {author} {\bibfnamefont {A.}~\bibnamefont
  {Tordesillas}}, \bibinfo {author} {\bibfnamefont {J.~E.}\ \bibnamefont
  {Hilton}},\ and\ \bibinfo {author} {\bibfnamefont {S.~T.}\ \bibnamefont
  {Tobin}},\ }\bibfield  {title} {\enquote {\bibinfo {title} {Stick-slip and
  force chain evolution in a granular bed in response to a grain intruder},}\
  }\href {https://doi.org/10.1103/PhysRevE.89.042207} {\bibfield  {journal}
  {\bibinfo  {journal} {Phys. Rev. E}\ }\textbf {\bibinfo {volume} {89}},\
  \bibinfo {pages} {042207} (\bibinfo {year} {2014})}\BibitemShut {NoStop}%
\bibitem [{\citenamefont {Kozlowski}\ \emph {et~al.}(2019)\citenamefont
  {Kozlowski}, \citenamefont {Carlevaro}, \citenamefont {Daniels},
  \citenamefont {Kondic}, \citenamefont {Pugnaloni}, \citenamefont {Socolar},
  \citenamefont {Zheng},\ and\ \citenamefont {Behringer}}]{Kozlowski}%
  \BibitemOpen
  \bibfield  {author} {\bibinfo {author} {\bibfnamefont {R.}~\bibnamefont
  {Kozlowski}}, \bibinfo {author} {\bibfnamefont {C.~M.}\ \bibnamefont
  {Carlevaro}}, \bibinfo {author} {\bibfnamefont {K.~E.}\ \bibnamefont
  {Daniels}}, \bibinfo {author} {\bibfnamefont {L.}~\bibnamefont {Kondic}},
  \bibinfo {author} {\bibfnamefont {L.~A.}\ \bibnamefont {Pugnaloni}}, \bibinfo
  {author} {\bibfnamefont {J.~E.~S.}\ \bibnamefont {Socolar}}, \bibinfo
  {author} {\bibfnamefont {H.}~\bibnamefont {Zheng}},\ and\ \bibinfo {author}
  {\bibfnamefont {R.~P.}\ \bibnamefont {Behringer}},\ }\bibfield  {title}
  {\enquote {\bibinfo {title} {Dynamics of a grain-scale intruder in a
  two-dimensional granular medium with and without basal friction},}\ }\href
  {https://doi.org/10.1103/PhysRevE.100.032905} {\bibfield  {journal} {\bibinfo
   {journal} {Phys. Rev. E}\ }\textbf {\bibinfo {volume} {100}},\ \bibinfo
  {pages} {032905} (\bibinfo {year} {2019})}\BibitemShut {NoStop}%
\bibitem [{\citenamefont {Carlevaro}\ \emph {et~al.}(2020)\citenamefont
  {Carlevaro}, \citenamefont {Kozlowski}, \citenamefont {Pugnaloni},
  \citenamefont {Zheng}, \citenamefont {Socolar},\ and\ \citenamefont
  {Kondic}}]{Carlevaro}%
  \BibitemOpen
  \bibfield  {author} {\bibinfo {author} {\bibfnamefont {C.~M.}\ \bibnamefont
  {Carlevaro}}, \bibinfo {author} {\bibfnamefont {R.}~\bibnamefont
  {Kozlowski}}, \bibinfo {author} {\bibfnamefont {L.~A.}\ \bibnamefont
  {Pugnaloni}}, \bibinfo {author} {\bibfnamefont {H.}~\bibnamefont {Zheng}},
  \bibinfo {author} {\bibfnamefont {J.~E.~S.}\ \bibnamefont {Socolar}},\ and\
  \bibinfo {author} {\bibfnamefont {L.}~\bibnamefont {Kondic}},\ }\bibfield
  {title} {\enquote {\bibinfo {title} {Intruder in a two-dimensional granular
  system: Effects of dynamic and static basal friction on stick-slip and
  clogging dynamics},}\ }\href {https://doi.org/10.1103/PhysRevE.101.012909}
  {\bibfield  {journal} {\bibinfo  {journal} {Phys. Rev. E}\ }\textbf {\bibinfo
  {volume} {101}},\ \bibinfo {pages} {012909} (\bibinfo {year}
  {2020})}\BibitemShut {NoStop}%
\bibitem [{\citenamefont {Kozlowski}\ \emph {et~al.}(2022)\citenamefont
  {Kozlowski}, \citenamefont {Zheng}, \citenamefont {Daniels},\ and\
  \citenamefont {Socolar}}]{Kozlowski2}%
  \BibitemOpen
  \bibfield  {author} {\bibinfo {author} {\bibfnamefont {R.}~\bibnamefont
  {Kozlowski}}, \bibinfo {author} {\bibfnamefont {H.}~\bibnamefont {Zheng}},
  \bibinfo {author} {\bibfnamefont {K.~E.}\ \bibnamefont {Daniels}},\ and\
  \bibinfo {author} {\bibfnamefont {J.~E.~S.}\ \bibnamefont {Socolar}},\
  }\bibfield  {title} {\enquote {\bibinfo {title} {Stick-slip dynamics in a
  granular material with varying grain angularity},}\ }\href
  {https://doi.org/10.3389/fphy.2022.916190} {\bibfield  {journal} {\bibinfo
  {journal} {Front. Phys.}\ }\textbf {\bibinfo {volume} {10}},\ \bibinfo
  {pages} {916190} (\bibinfo {year} {2022})}\BibitemShut {NoStop}%
\bibitem [{\citenamefont {Kozlowski}\ \emph {et~al.}(2021)\citenamefont
  {Kozlowski}, \citenamefont {Zheng}, \citenamefont {Daniels},\ and\
  \citenamefont {Socolar}}]{Kozlowski3}%
  \BibitemOpen
  \bibfield  {author} {\bibinfo {author} {\bibfnamefont {R.}~\bibnamefont
  {Kozlowski}}, \bibinfo {author} {\bibfnamefont {H.}~\bibnamefont {Zheng}},
  \bibinfo {author} {\bibfnamefont {K.~E.}\ \bibnamefont {Daniels}},\ and\
  \bibinfo {author} {\bibfnamefont {J.~E.~S.}\ \bibnamefont {Socolar}},\
  }\bibfield  {title} {\enquote {\bibinfo {title} {Stress propagation in
  locally loaded packings of disks and pentagons},}\ }\href
  {https://doi.org/10.1039/D1SM01137E} {\bibfield  {journal} {\bibinfo
  {journal} {Soft Matter}\ }\textbf {\bibinfo {volume} {17}},\ \bibinfo {pages}
  {10120--10127} (\bibinfo {year} {2021})}\BibitemShut {NoStop}%
\bibitem [{\citenamefont {Pugnaloni}\ \emph {et~al.}(2022)\citenamefont
  {Pugnaloni}, \citenamefont {Carlevaro}, \citenamefont {Kozlowski},
  \citenamefont {Zheng}, \citenamefont {Kondic},\ and\ \citenamefont
  {Socolar}}]{Pugnaloni}%
  \BibitemOpen
  \bibfield  {author} {\bibinfo {author} {\bibfnamefont {L.~A.}\ \bibnamefont
  {Pugnaloni}}, \bibinfo {author} {\bibfnamefont {C.~M.}\ \bibnamefont
  {Carlevaro}}, \bibinfo {author} {\bibfnamefont {R.}~\bibnamefont
  {Kozlowski}}, \bibinfo {author} {\bibfnamefont {H.}~\bibnamefont {Zheng}},
  \bibinfo {author} {\bibfnamefont {L.}~\bibnamefont {Kondic}},\ and\ \bibinfo
  {author} {\bibfnamefont {J.~E.~S.}\ \bibnamefont {Socolar}},\ }\bibfield
  {title} {\enquote {\bibinfo {title} {Universal features of the stick-slip
  dynamics of an intruder moving through a confined granular medium},}\ }\href
  {https://doi.org/10.1103/PhysRevE.105.L042902} {\bibfield  {journal}
  {\bibinfo  {journal} {Phys. Rev. E}\ }\textbf {\bibinfo {volume} {105}},\
  \bibinfo {pages} {L042902} (\bibinfo {year} {2022})}\BibitemShut {NoStop}%
\bibitem [{\citenamefont {Carvalho}, \citenamefont {Lima},\ and\ \citenamefont
  {Franklin}(2022)}]{Carvalho}%
  \BibitemOpen
  \bibfield  {author} {\bibinfo {author} {\bibfnamefont {D.~D.}\ \bibnamefont
  {Carvalho}}, \bibinfo {author} {\bibfnamefont {N.~C.}\ \bibnamefont {Lima}},\
  and\ \bibinfo {author} {\bibfnamefont {E.~M.}\ \bibnamefont {Franklin}},\
  }\bibfield  {title} {\enquote {\bibinfo {title} {Contacts, motion, and chain
  breaking in a two-dimensional granular system displaced by an intruder},}\
  }\href {https://doi.org/10.1103/PhysRevE.105.034903} {\bibfield  {journal}
  {\bibinfo  {journal} {Phys. Rev. E}\ }\textbf {\bibinfo {volume} {105}},\
  \bibinfo {pages} {034903} (\bibinfo {year} {2022})}\BibitemShut {NoStop}%
\bibitem [{\citenamefont {Andreotti}, \citenamefont {Forterre},\ and\
  \citenamefont {Pouliquen}(2013)}]{Andreotti_6}%
  \BibitemOpen
  \bibfield  {author} {\bibinfo {author} {\bibfnamefont {B.}~\bibnamefont
  {Andreotti}}, \bibinfo {author} {\bibfnamefont {Y.}~\bibnamefont
  {Forterre}},\ and\ \bibinfo {author} {\bibfnamefont {O.}~\bibnamefont
  {Pouliquen}},\ }\href@noop {} {\emph {\bibinfo {title} {Granular Media:
  {B}etween Fluid and Solid}}}\ (\bibinfo  {publisher} {Cambridge University
  Press},\ \bibinfo {year} {2013})\BibitemShut {NoStop}%
\bibitem [{\citenamefont {Radjai}\ \emph {et~al.}(1998)\citenamefont {Radjai},
  \citenamefont {Wolf}, \citenamefont {Jean},\ and\ \citenamefont
  {Moreau}}]{Radjai1}%
  \BibitemOpen
  \bibfield  {author} {\bibinfo {author} {\bibfnamefont {F.}~\bibnamefont
  {Radjai}}, \bibinfo {author} {\bibfnamefont {D.~E.}\ \bibnamefont {Wolf}},
  \bibinfo {author} {\bibfnamefont {M.}~\bibnamefont {Jean}},\ and\ \bibinfo
  {author} {\bibfnamefont {J.-J.}\ \bibnamefont {Moreau}},\ }\bibfield  {title}
  {\enquote {\bibinfo {title} {Bimodal character of stress transmission in
  granular packings},}\ }\href {https://doi.org/10.1103/PhysRevLett.80.61}
  {\bibfield  {journal} {\bibinfo  {journal} {Phys. Rev. Lett.}\ }\textbf
  {\bibinfo {volume} {80}},\ \bibinfo {pages} {61--64} (\bibinfo {year}
  {1998})}\BibitemShut {NoStop}%
\bibitem [{\citenamefont {Majmudar}\ and\ \citenamefont
  {Behringer}(2005)}]{Majmudar}%
  \BibitemOpen
  \bibfield  {author} {\bibinfo {author} {\bibfnamefont {T.~S.}\ \bibnamefont
  {Majmudar}}\ and\ \bibinfo {author} {\bibfnamefont {R.~P.}\ \bibnamefont
  {Behringer}},\ }\bibfield  {title} {\enquote {\bibinfo {title} {Contact force
  measurements and stress-induced anisotropy in granular materials},}\
  }\href@noop {} {\bibfield  {journal} {\bibinfo  {journal} {Nature}\ }\textbf
  {\bibinfo {volume} {435}},\ \bibinfo {pages} {1079--1082} (\bibinfo {year}
  {2005})}\BibitemShut {NoStop}%
\bibitem [{\citenamefont {Cates}\ \emph {et~al.}(1998)\citenamefont {Cates},
  \citenamefont {Wittmer}, \citenamefont {Bouchaud},\ and\ \citenamefont
  {Claudin}}]{Cates}%
  \BibitemOpen
  \bibfield  {author} {\bibinfo {author} {\bibfnamefont {M.~E.}\ \bibnamefont
  {Cates}}, \bibinfo {author} {\bibfnamefont {J.~P.}\ \bibnamefont {Wittmer}},
  \bibinfo {author} {\bibfnamefont {J.-P.}\ \bibnamefont {Bouchaud}},\ and\
  \bibinfo {author} {\bibfnamefont {P.}~\bibnamefont {Claudin}},\ }\bibfield
  {title} {\enquote {\bibinfo {title} {Jamming, force chains, and fragile
  matter},}\ }\href {https://doi.org/10.1103/PhysRevLett.81.1841} {\bibfield
  {journal} {\bibinfo  {journal} {Phys. Rev. Lett.}\ }\textbf {\bibinfo
  {volume} {81}},\ \bibinfo {pages} {1841--1844} (\bibinfo {year}
  {1998})}\BibitemShut {NoStop}%
\bibitem [{\citenamefont {Bi}\ \emph {et~al.}(2011)\citenamefont {Bi},
  \citenamefont {Zhang}, \citenamefont {Chakraborty},\ and\ \citenamefont
  {Behringer}}]{Bi}%
  \BibitemOpen
  \bibfield  {author} {\bibinfo {author} {\bibfnamefont {D.}~\bibnamefont
  {Bi}}, \bibinfo {author} {\bibfnamefont {J.}~\bibnamefont {Zhang}}, \bibinfo
  {author} {\bibfnamefont {B.}~\bibnamefont {Chakraborty}},\ and\ \bibinfo
  {author} {\bibfnamefont {R.~P.}\ \bibnamefont {Behringer}},\ }\bibfield
  {title} {\enquote {\bibinfo {title} {Jamming by shear},}\ }\href@noop {}
  {\bibfield  {journal} {\bibinfo  {journal} {Nature}\ }\textbf {\bibinfo
  {volume} {480}},\ \bibinfo {pages} {355--358} (\bibinfo {year}
  {2011})}\BibitemShut {NoStop}%
\bibitem [{\citenamefont {Seguin}(2020)}]{Seguin2}%
  \BibitemOpen
  \bibfield  {author} {\bibinfo {author} {\bibfnamefont {A.}~\bibnamefont
  {Seguin}},\ }\bibfield  {title} {\enquote {\bibinfo {title} {Experimental
  study of some properties of the strong and weak force networks in a jammed
  granular medium},}\ }\href {https://doi.org/10.1007/s10035-020-01015-z}
  {\bibfield  {journal} {\bibinfo  {journal} {Granular Matter}\ }\textbf
  {\bibinfo {volume} {22}} (\bibinfo {year} {2020}),\
  10.1007/s10035-020-01015-z}\BibitemShut {NoStop}%
\bibitem [{\citenamefont {Behringer}\ and\ \citenamefont
  {Chakraborty}(2018)}]{Behringer_1}%
  \BibitemOpen
  \bibfield  {author} {\bibinfo {author} {\bibfnamefont {R.~P.}\ \bibnamefont
  {Behringer}}\ and\ \bibinfo {author} {\bibfnamefont {B.}~\bibnamefont
  {Chakraborty}},\ }\bibfield  {title} {\enquote {\bibinfo {title} {The physics
  of jamming for granular materials: {A} review},}\ }\href@noop {} {\bibfield
  {journal} {\bibinfo  {journal} {Rep. Prog. Phys.}\ }\textbf {\bibinfo
  {volume} {82}},\ \bibinfo {pages} {012601} (\bibinfo {year}
  {2018})}\BibitemShut {NoStop}%
\bibitem [{\citenamefont {Featherstone}\ \emph {et~al.}(2021)\citenamefont
  {Featherstone}, \citenamefont {Bullard}, \citenamefont {Emm}, \citenamefont
  {Jackson}, \citenamefont {Reid}, \citenamefont {Shefferman}, \citenamefont
  {Dove}, \citenamefont {Colwell}, \citenamefont {Kollmer},\ and\ \citenamefont
  {Daniels}}]{Featherstone}%
  \BibitemOpen
  \bibfield  {author} {\bibinfo {author} {\bibfnamefont {J.}~\bibnamefont
  {Featherstone}}, \bibinfo {author} {\bibfnamefont {R.}~\bibnamefont
  {Bullard}}, \bibinfo {author} {\bibfnamefont {T.}~\bibnamefont {Emm}},
  \bibinfo {author} {\bibfnamefont {A.}~\bibnamefont {Jackson}}, \bibinfo
  {author} {\bibfnamefont {R.}~\bibnamefont {Reid}}, \bibinfo {author}
  {\bibfnamefont {S.}~\bibnamefont {Shefferman}}, \bibinfo {author}
  {\bibfnamefont {A.}~\bibnamefont {Dove}}, \bibinfo {author} {\bibfnamefont
  {J.}~\bibnamefont {Colwell}}, \bibinfo {author} {\bibfnamefont {J.~E.}\
  \bibnamefont {Kollmer}},\ and\ \bibinfo {author} {\bibfnamefont {K.~E.}\
  \bibnamefont {Daniels}},\ }\bibfield  {title} {\enquote {\bibinfo {title}
  {Stick-slip dynamics in penetration experiments on simulated regolith},}\
  }\href {https://doi.org/10.3847/psj/ac3de2} {\bibfield  {journal} {\bibinfo
  {journal} {The Planetary Science Journal}\ }\textbf {\bibinfo {volume} {2}},\
  \bibinfo {pages} {243} (\bibinfo {year} {2021})}\BibitemShut {NoStop}%
\bibitem [{\citenamefont {Seguin}\ \emph {et~al.}(2016)\citenamefont {Seguin},
  \citenamefont {Coulais}, \citenamefont {Martinez}, \citenamefont {Bertho},\
  and\ \citenamefont {Gondret}}]{Seguin1}%
  \BibitemOpen
  \bibfield  {author} {\bibinfo {author} {\bibfnamefont {A.}~\bibnamefont
  {Seguin}}, \bibinfo {author} {\bibfnamefont {C.}~\bibnamefont {Coulais}},
  \bibinfo {author} {\bibfnamefont {F.}~\bibnamefont {Martinez}}, \bibinfo
  {author} {\bibfnamefont {Y.}~\bibnamefont {Bertho}},\ and\ \bibinfo {author}
  {\bibfnamefont {P.}~\bibnamefont {Gondret}},\ }\bibfield  {title} {\enquote
  {\bibinfo {title} {Local rheological measurements in the granular flow around
  an intruder},}\ }\href {https://doi.org/10.1103/PhysRevE.93.012904}
  {\bibfield  {journal} {\bibinfo  {journal} {Phys. Rev. E}\ }\textbf {\bibinfo
  {volume} {93}},\ \bibinfo {pages} {012904} (\bibinfo {year}
  {2016})}\BibitemShut {NoStop}%
\bibitem [{\citenamefont {Albert}\ \emph {et~al.}(1999)\citenamefont {Albert},
  \citenamefont {Pfeifer}, \citenamefont {Barab\'asi},\ and\ \citenamefont
  {Schiffer}}]{Albert}%
  \BibitemOpen
  \bibfield  {author} {\bibinfo {author} {\bibfnamefont {R.}~\bibnamefont
  {Albert}}, \bibinfo {author} {\bibfnamefont {M.~A.}\ \bibnamefont {Pfeifer}},
  \bibinfo {author} {\bibfnamefont {A.-L.}\ \bibnamefont {Barab\'asi}},\ and\
  \bibinfo {author} {\bibfnamefont {P.}~\bibnamefont {Schiffer}},\ }\bibfield
  {title} {\enquote {\bibinfo {title} {Slow drag in a granular medium},}\
  }\href {https://doi.org/10.1103/PhysRevLett.82.205} {\bibfield  {journal}
  {\bibinfo  {journal} {Phys. Rev. Lett.}\ }\textbf {\bibinfo {volume} {82}},\
  \bibinfo {pages} {205--208} (\bibinfo {year} {1999})}\BibitemShut {NoStop}%
\bibitem [{\citenamefont {Albert}\ \emph {et~al.}(2001)\citenamefont {Albert},
  \citenamefont {Sample}, \citenamefont {Morss}, \citenamefont {Rajagopalan},
  \citenamefont {Barab\'asi},\ and\ \citenamefont {Schiffer}}]{Albert2}%
  \BibitemOpen
  \bibfield  {author} {\bibinfo {author} {\bibfnamefont {I.}~\bibnamefont
  {Albert}}, \bibinfo {author} {\bibfnamefont {J.~G.}\ \bibnamefont {Sample}},
  \bibinfo {author} {\bibfnamefont {A.~J.}\ \bibnamefont {Morss}}, \bibinfo
  {author} {\bibfnamefont {S.}~\bibnamefont {Rajagopalan}}, \bibinfo {author}
  {\bibfnamefont {A.-L.}\ \bibnamefont {Barab\'asi}},\ and\ \bibinfo {author}
  {\bibfnamefont {P.}~\bibnamefont {Schiffer}},\ }\bibfield  {title} {\enquote
  {\bibinfo {title} {Granular drag on a discrete object: Shape effects on
  jamming},}\ }\href {https://doi.org/10.1103/PhysRevE.64.061303} {\bibfield
  {journal} {\bibinfo  {journal} {Phys. Rev. E}\ }\textbf {\bibinfo {volume}
  {64}},\ \bibinfo {pages} {061303} (\bibinfo {year} {2001})}\BibitemShut
  {NoStop}%
\bibitem [{\citenamefont {Stone}\ \emph {et~al.}(2004)\citenamefont {Stone},
  \citenamefont {Barry}, \citenamefont {Bernstein}, \citenamefont {Pelc},
  \citenamefont {Tsui},\ and\ \citenamefont {Schiffer}}]{Stone}%
  \BibitemOpen
  \bibfield  {author} {\bibinfo {author} {\bibfnamefont {M.~B.}\ \bibnamefont
  {Stone}}, \bibinfo {author} {\bibfnamefont {R.}~\bibnamefont {Barry}},
  \bibinfo {author} {\bibfnamefont {D.~P.}\ \bibnamefont {Bernstein}}, \bibinfo
  {author} {\bibfnamefont {M.~D.}\ \bibnamefont {Pelc}}, \bibinfo {author}
  {\bibfnamefont {Y.~K.}\ \bibnamefont {Tsui}},\ and\ \bibinfo {author}
  {\bibfnamefont {P.}~\bibnamefont {Schiffer}},\ }\bibfield  {title} {\enquote
  {\bibinfo {title} {Local jamming via penetration of a granular medium},}\
  }\href {https://doi.org/10.1103/PhysRevE.70.041301} {\bibfield  {journal}
  {\bibinfo  {journal} {Phys. Rev. E}\ }\textbf {\bibinfo {volume} {70}},\
  \bibinfo {pages} {041301} (\bibinfo {year} {2004})}\BibitemShut {NoStop}%
\bibitem [{\citenamefont {Geng}\ and\ \citenamefont {Behringer}(2005)}]{Geng}%
  \BibitemOpen
  \bibfield  {author} {\bibinfo {author} {\bibfnamefont {J.}~\bibnamefont
  {Geng}}\ and\ \bibinfo {author} {\bibfnamefont {R.~P.}\ \bibnamefont
  {Behringer}},\ }\bibfield  {title} {\enquote {\bibinfo {title} {Slow drag in
  two-dimensional granular media},}\ }\href
  {https://doi.org/10.1103/PhysRevE.71.011302} {\bibfield  {journal} {\bibinfo
  {journal} {Phys. Rev. E}\ }\textbf {\bibinfo {volume} {71}},\ \bibinfo
  {pages} {011302} (\bibinfo {year} {2005})}\BibitemShut {NoStop}%
\bibitem [{\citenamefont {Costantino}\ \emph {et~al.}(2008)\citenamefont
  {Costantino}, \citenamefont {Scheidemantel}, \citenamefont {Stone},
  \citenamefont {Conger}, \citenamefont {Klein}, \citenamefont {Lohr},
  \citenamefont {Modig},\ and\ \citenamefont {Schiffer}}]{Costantino}%
  \BibitemOpen
  \bibfield  {author} {\bibinfo {author} {\bibfnamefont {D.~J.}\ \bibnamefont
  {Costantino}}, \bibinfo {author} {\bibfnamefont {T.~J.}\ \bibnamefont
  {Scheidemantel}}, \bibinfo {author} {\bibfnamefont {M.~B.}\ \bibnamefont
  {Stone}}, \bibinfo {author} {\bibfnamefont {C.}~\bibnamefont {Conger}},
  \bibinfo {author} {\bibfnamefont {K.}~\bibnamefont {Klein}}, \bibinfo
  {author} {\bibfnamefont {M.}~\bibnamefont {Lohr}}, \bibinfo {author}
  {\bibfnamefont {Z.}~\bibnamefont {Modig}},\ and\ \bibinfo {author}
  {\bibfnamefont {P.}~\bibnamefont {Schiffer}},\ }\bibfield  {title} {\enquote
  {\bibinfo {title} {Starting to move through a granular medium},}\ }\href
  {https://doi.org/10.1103/PhysRevLett.101.108001} {\bibfield  {journal}
  {\bibinfo  {journal} {Phys. Rev. Lett.}\ }\textbf {\bibinfo {volume} {101}},\
  \bibinfo {pages} {108001} (\bibinfo {year} {2008})}\BibitemShut {NoStop}%
\bibitem [{\citenamefont {Tripura}\ \emph {et~al.}(2022)\citenamefont
  {Tripura}, \citenamefont {Kumar}, \citenamefont {Anyam},\ and\ \citenamefont
  {Reddy}}]{Tripura}%
  \BibitemOpen
  \bibfield  {author} {\bibinfo {author} {\bibfnamefont {B.~K.}\ \bibnamefont
  {Tripura}}, \bibinfo {author} {\bibfnamefont {S.}~\bibnamefont {Kumar}},
  \bibinfo {author} {\bibfnamefont {V.~K.~R.}\ \bibnamefont {Anyam}},\ and\
  \bibinfo {author} {\bibfnamefont {K.~A.}\ \bibnamefont {Reddy}},\ }\bibfield
  {title} {\enquote {\bibinfo {title} {Drag on a circular intruder traversing a
  shape-heterogeneous granular mixture},}\ }\href
  {https://doi.org/10.1103/PhysRevE.106.014901} {\bibfield  {journal} {\bibinfo
   {journal} {Phys. Rev. E}\ }\textbf {\bibinfo {volume} {106}},\ \bibinfo
  {pages} {014901} (\bibinfo {year} {2022})}\BibitemShut {NoStop}%
\bibitem [{\citenamefont {Pacheco-V\'azquez}\ and\ \citenamefont
  {Ruiz-Su\'arez}(2010)}]{Pacheco}%
  \BibitemOpen
  \bibfield  {author} {\bibinfo {author} {\bibfnamefont {F.}~\bibnamefont
  {Pacheco-V\'azquez}}\ and\ \bibinfo {author} {\bibfnamefont {J.}~\bibnamefont
  {Ruiz-Su\'arez}},\ }\bibfield  {title} {\enquote {\bibinfo {title}
  {Cooperative dynamics in the penetration of a group of intruders in a
  granular medium},}\ }\href@noop {} {\bibfield  {journal} {\bibinfo  {journal}
  {Nat. Commun.}\ }\textbf {\bibinfo {volume} {1}} (\bibinfo {year}
  {2010})}\BibitemShut {NoStop}%
\bibitem [{\citenamefont {Solano-Altamirano}\ \emph {et~al.}(2013)\citenamefont
  {Solano-Altamirano}, \citenamefont {Caballero-Robledo}, \citenamefont
  {Pacheco-V\'azquez}, \citenamefont {Kamphorst},\ and\ \citenamefont
  {Ruiz-Su\'arez}}]{Solano}%
  \BibitemOpen
  \bibfield  {author} {\bibinfo {author} {\bibfnamefont {J.~M.}\ \bibnamefont
  {Solano-Altamirano}}, \bibinfo {author} {\bibfnamefont {G.~A.}\ \bibnamefont
  {Caballero-Robledo}}, \bibinfo {author} {\bibfnamefont {F.}~\bibnamefont
  {Pacheco-V\'azquez}}, \bibinfo {author} {\bibfnamefont {V.}~\bibnamefont
  {Kamphorst}},\ and\ \bibinfo {author} {\bibfnamefont {J.~C.}\ \bibnamefont
  {Ruiz-Su\'arez}},\ }\bibfield  {title} {\enquote {\bibinfo {title}
  {Flow-mediated coupling on projectiles falling within a superlight granular
  medium},}\ }\href {https://doi.org/10.1103/PhysRevE.88.032206} {\bibfield
  {journal} {\bibinfo  {journal} {Phys. Rev. E}\ }\textbf {\bibinfo {volume}
  {88}},\ \bibinfo {pages} {032206} (\bibinfo {year} {2013})}\BibitemShut
  {NoStop}%
\bibitem [{\citenamefont {Dhiman}\ \emph {et~al.}(2020)\citenamefont {Dhiman},
  \citenamefont {Kumar}, \citenamefont {Anki~Reddy},\ and\ \citenamefont
  {Gupta}}]{Dhiman}%
  \BibitemOpen
  \bibfield  {author} {\bibinfo {author} {\bibfnamefont {M.}~\bibnamefont
  {Dhiman}}, \bibinfo {author} {\bibfnamefont {S.}~\bibnamefont {Kumar}},
  \bibinfo {author} {\bibfnamefont {K.}~\bibnamefont {Anki~Reddy}},\ and\
  \bibinfo {author} {\bibfnamefont {R.}~\bibnamefont {Gupta}},\ }\bibfield
  {title} {\enquote {\bibinfo {title} {Origin of the long-ranged attraction or
  repulsion between intruders in a confined granular medium},}\ }\href@noop {}
  {\bibfield  {journal} {\bibinfo  {journal} {J. Fluid Mech.}\ }\textbf
  {\bibinfo {volume} {886}},\ \bibinfo {pages} {A23} (\bibinfo {year}
  {2020})}\BibitemShut {NoStop}%
\bibitem [{\citenamefont {Kawabata}\ \emph {et~al.}(2020)\citenamefont
  {Kawabata}, \citenamefont {Yoshida}, \citenamefont {Shimosaka},\ and\
  \citenamefont {Shirakawa}}]{Kawabata}%
  \BibitemOpen
  \bibfield  {author} {\bibinfo {author} {\bibfnamefont {D.}~\bibnamefont
  {Kawabata}}, \bibinfo {author} {\bibfnamefont {M.}~\bibnamefont {Yoshida}},
  \bibinfo {author} {\bibfnamefont {A.}~\bibnamefont {Shimosaka}},\ and\
  \bibinfo {author} {\bibfnamefont {Y.}~\bibnamefont {Shirakawa}},\ }\bibfield
  {title} {\enquote {\bibinfo {title} {Discrete element method simulation
  analysis of the generation mechanism of cooperative behavior of disks falling
  in a low-density particle bed},}\ }\href@noop {} {\bibfield  {journal}
  {\bibinfo  {journal} {Adv. Powder Technol.}\ }\textbf {\bibinfo {volume}
  {31}},\ \bibinfo {pages} {1381--1390} (\bibinfo {year} {2020})}\BibitemShut
  {NoStop}%
\bibitem [{\citenamefont {Pravin}\ \emph {et~al.}(2021)\citenamefont {Pravin},
  \citenamefont {Chang}, \citenamefont {Han}, \citenamefont {London},
  \citenamefont {Goldman}, \citenamefont {Jaeger},\ and\ \citenamefont
  {Hsieh}}]{Pravin}%
  \BibitemOpen
  \bibfield  {author} {\bibinfo {author} {\bibfnamefont {S.}~\bibnamefont
  {Pravin}}, \bibinfo {author} {\bibfnamefont {B.}~\bibnamefont {Chang}},
  \bibinfo {author} {\bibfnamefont {E.}~\bibnamefont {Han}}, \bibinfo {author}
  {\bibfnamefont {L.}~\bibnamefont {London}}, \bibinfo {author} {\bibfnamefont
  {D.~I.}\ \bibnamefont {Goldman}}, \bibinfo {author} {\bibfnamefont {H.~M.}\
  \bibnamefont {Jaeger}},\ and\ \bibinfo {author} {\bibfnamefont {S.~T.}\
  \bibnamefont {Hsieh}},\ }\bibfield  {title} {\enquote {\bibinfo {title}
  {Effect of two parallel intruders on total work during granular
  penetrations},}\ }\href {https://doi.org/10.1103/PhysRevE.104.024902}
  {\bibfield  {journal} {\bibinfo  {journal} {Phys. Rev. E}\ }\textbf {\bibinfo
  {volume} {104}},\ \bibinfo {pages} {024902} (\bibinfo {year}
  {2021})}\BibitemShut {NoStop}%
\bibitem [{\citenamefont {Espinosa}\ \emph {et~al.}(2022)\citenamefont
  {Espinosa}, \citenamefont {D\'iaz-Meli\'an}, \citenamefont
  {Serrano-Mu\~noz},\ and\ \citenamefont {Altshuler}}]{Espinosa}%
  \BibitemOpen
  \bibfield  {author} {\bibinfo {author} {\bibfnamefont {M.}~\bibnamefont
  {Espinosa}}, \bibinfo {author} {\bibfnamefont {V.}~\bibnamefont
  {D\'iaz-Meli\'an}}, \bibinfo {author} {\bibfnamefont {A.}~\bibnamefont
  {Serrano-Mu\~noz}},\ and\ \bibinfo {author} {\bibfnamefont {E.}~\bibnamefont
  {Altshuler}},\ }\bibfield  {title} {\enquote {\bibinfo {title} {Intruders
  cooperatively interact with a wall into granular matter},}\ }\href
  {https://doi.org/10.1007/s10035-021-01200-8} {\bibfield  {journal} {\bibinfo
  {journal} {Granular Matter}\ }\textbf {\bibinfo {volume} {24}} (\bibinfo
  {year} {2022}),\ 10.1007/s10035-021-01200-8}\BibitemShut {NoStop}%
\bibitem [{\citenamefont {Merceron}, \citenamefont {Sauret},\ and\
  \citenamefont {Jop}(2018)}]{Merceron}%
  \BibitemOpen
  \bibfield  {author} {\bibinfo {author} {\bibfnamefont {A.}~\bibnamefont
  {Merceron}}, \bibinfo {author} {\bibfnamefont {A.}~\bibnamefont {Sauret}},\
  and\ \bibinfo {author} {\bibfnamefont {P.}~\bibnamefont {Jop}},\ }\bibfield
  {title} {\enquote {\bibinfo {title} {Cooperative effects induced by intruders
  evolving through a granular medium},}\ }\href@noop {} {\bibfield  {journal}
  {\bibinfo  {journal} {{EPL}-Europhys. Lett.}\ }\textbf {\bibinfo {volume}
  {121}},\ \bibinfo {pages} {34005} (\bibinfo {year} {2018})}\BibitemShut
  {NoStop}%
\bibitem [{\citenamefont {Kloss}\ and\ \citenamefont {Goniva}(2010)}]{Kloss}%
  \BibitemOpen
  \bibfield  {author} {\bibinfo {author} {\bibfnamefont {C.}~\bibnamefont
  {Kloss}}\ and\ \bibinfo {author} {\bibfnamefont {C.}~\bibnamefont {Goniva}},\
  }\bibfield  {title} {\enquote {\bibinfo {title} {{LIGGGHTS}: a new open
  source discrete element simulation software},}\ }in\ \href@noop {} {\emph
  {\bibinfo {booktitle} {Proc. 5th Int. Conf. on Discrete Element Methods}}}\
  (\bibinfo {address} {London, UK},\ \bibinfo {year} {2010})\BibitemShut
  {NoStop}%
\bibitem [{\citenamefont {Berger}\ \emph {et~al.}(2015)\citenamefont {Berger},
  \citenamefont {Kloss}, \citenamefont {Kohlmeyer},\ and\ \citenamefont
  {Pirker}}]{Berger}%
  \BibitemOpen
  \bibfield  {author} {\bibinfo {author} {\bibfnamefont {R.}~\bibnamefont
  {Berger}}, \bibinfo {author} {\bibfnamefont {C.}~\bibnamefont {Kloss}},
  \bibinfo {author} {\bibfnamefont {A.}~\bibnamefont {Kohlmeyer}},\ and\
  \bibinfo {author} {\bibfnamefont {S.}~\bibnamefont {Pirker}},\ }\bibfield
  {title} {\enquote {\bibinfo {title} {Hybrid parallelization of the {LIGGGHTS}
  open-source {DEM} code},}\ }\href@noop {} {\bibfield  {journal} {\bibinfo
  {journal} {Powder Technol.}\ }\textbf {\bibinfo {volume} {278}},\ \bibinfo
  {pages} {234--247} (\bibinfo {year} {2015})}\BibitemShut {NoStop}%
\bibitem [{\citenamefont {Herman}(2016)}]{Herman}%
  \BibitemOpen
  \bibfield  {author} {\bibinfo {author} {\bibfnamefont {A.}~\bibnamefont
  {Herman}},\ }\bibfield  {title} {\enquote {\bibinfo {title}
  {Discrete-{E}lement bonded-particle {S}ea {I}ce model {DESI}gn, version 1.3
  a--model description and implementation},}\ }\href
  {https://doi.org/10.5194/gmd-9-1219-2016} {\bibfield  {journal} {\bibinfo
  {journal} {Geosci. Model Dev.}\ }\textbf {\bibinfo {volume} {9}},\ \bibinfo
  {pages} {1219--1241} (\bibinfo {year} {2016})}\BibitemShut {NoStop}%
\bibitem [{\citenamefont {Cundall}\ and\ \citenamefont
  {Strack}(1979)}]{Cundall}%
  \BibitemOpen
  \bibfield  {author} {\bibinfo {author} {\bibfnamefont {P.~A.}\ \bibnamefont
  {Cundall}}\ and\ \bibinfo {author} {\bibfnamefont {O.~D.}\ \bibnamefont
  {Strack}},\ }\bibfield  {title} {\enquote {\bibinfo {title} {A discrete
  numerical model for granular assemblies},}\ }\href@noop {} {\bibfield
  {journal} {\bibinfo  {journal} {G\'eotechnique}\ }\textbf {\bibinfo {volume}
  {29}},\ \bibinfo {pages} {47--65} (\bibinfo {year} {1979})}\BibitemShut
  {NoStop}%
\bibitem [{\citenamefont {{Di Renzo}}\ and\ \citenamefont {{Di
  Maio}}(2005)}]{direnzo}%
  \BibitemOpen
  \bibfield  {author} {\bibinfo {author} {\bibfnamefont {A.}~\bibnamefont {{Di
  Renzo}}}\ and\ \bibinfo {author} {\bibfnamefont {F.~P.}\ \bibnamefont {{Di
  Maio}}},\ }\bibfield  {title} {\enquote {\bibinfo {title} {An improved
  integral non-linear model for the contact of particles in distinct element
  simulations},}\ }\href@noop {} {\bibfield  {journal} {\bibinfo  {journal}
  {Chem. Eng. Sci.}\ }\textbf {\bibinfo {volume} {60}},\ \bibinfo {pages}
  {1303--1312} (\bibinfo {year} {2005})}\BibitemShut {NoStop}%
\bibitem [{\citenamefont {Speedy}(1999)}]{Speedy}%
  \BibitemOpen
  \bibfield  {author} {\bibinfo {author} {\bibfnamefont {R.~J.}\ \bibnamefont
  {Speedy}},\ }\bibfield  {title} {\enquote {\bibinfo {title} {Glass transition
  in hard disc mixtures},}\ }\href@noop {} {\bibfield  {journal} {\bibinfo
  {journal} {J. Chem. Phys.}\ }\textbf {\bibinfo {volume} {110}},\ \bibinfo
  {pages} {4559--4565} (\bibinfo {year} {1999})}\BibitemShut {NoStop}%
\bibitem [{\citenamefont {Hashemnia}\ and\ \citenamefont
  {Spelt}(2014)}]{Hashemnia}%
  \BibitemOpen
  \bibfield  {author} {\bibinfo {author} {\bibfnamefont {K.}~\bibnamefont
  {Hashemnia}}\ and\ \bibinfo {author} {\bibfnamefont {J.~K.}\ \bibnamefont
  {Spelt}},\ }\bibfield  {title} {\enquote {\bibinfo {title} {Particle impact
  velocities in a vibrationally fluidized granular flow: measurements and
  discrete element predictions},}\ }\href
  {https://doi.org/https://doi.org/10.1016/j.ces.2014.01.027} {\bibfield
  {journal} {\bibinfo  {journal} {Chem. Eng. Sci.}\ }\textbf {\bibinfo {volume}
  {109}},\ \bibinfo {pages} {123--135} (\bibinfo {year} {2014})}\BibitemShut
  {NoStop}%
\bibitem [{\citenamefont {Gondret}, \citenamefont {Lance},\ and\ \citenamefont
  {Petit}(2002)}]{Gondret}%
  \BibitemOpen
  \bibfield  {author} {\bibinfo {author} {\bibfnamefont {P.}~\bibnamefont
  {Gondret}}, \bibinfo {author} {\bibfnamefont {M.}~\bibnamefont {Lance}},\
  and\ \bibinfo {author} {\bibfnamefont {L.}~\bibnamefont {Petit}},\ }\bibfield
   {title} {\enquote {\bibinfo {title} {Bouncing motion of spherical particles
  in fluids},}\ }\href {https://doi.org/10.1063/1.1427920} {\bibfield
  {journal} {\bibinfo  {journal} {Phys. Fluids}\ }\textbf {\bibinfo {volume}
  {14}},\ \bibinfo {pages} {643--652} (\bibinfo {year} {2002})},\ \Eprint
  {https://arxiv.org/abs/https://doi.org/10.1063/1.1427920}
  {https://doi.org/10.1063/1.1427920} \BibitemShut {NoStop}%
\bibitem [{\citenamefont {Zaikin}\ \emph {et~al.}(2017)\citenamefont {Zaikin},
  \citenamefont {Korablin}, \citenamefont {Dyulger},\ and\ \citenamefont
  {Barnenkov}}]{Zaikin}%
  \BibitemOpen
  \bibfield  {author} {\bibinfo {author} {\bibfnamefont {O.}~\bibnamefont
  {Zaikin}}, \bibinfo {author} {\bibfnamefont {A.}~\bibnamefont {Korablin}},
  \bibinfo {author} {\bibfnamefont {N.}~\bibnamefont {Dyulger}},\ and\ \bibinfo
  {author} {\bibfnamefont {N.}~\bibnamefont {Barnenkov}},\ }\bibfield  {title}
  {\enquote {\bibinfo {title} {Model of the relationship between the velocity
  restitution coefficient and the initial car velocity during collision},}\
  }\href@noop {} {\bibfield  {journal} {\bibinfo  {journal} {Transp. Res.
  Proc.}\ }\textbf {\bibinfo {volume} {20}},\ \bibinfo {pages} {717--723}
  (\bibinfo {year} {2017})}\BibitemShut {NoStop}%
\bibitem [{\citenamefont {Lommen}, \citenamefont {Schott},\ and\ \citenamefont
  {Lodewijks}(2014)}]{Lommen}%
  \BibitemOpen
  \bibfield  {author} {\bibinfo {author} {\bibfnamefont {S.}~\bibnamefont
  {Lommen}}, \bibinfo {author} {\bibfnamefont {D.}~\bibnamefont {Schott}},\
  and\ \bibinfo {author} {\bibfnamefont {G.}~\bibnamefont {Lodewijks}},\
  }\bibfield  {title} {\enquote {\bibinfo {title} {{DEM} speedup: {S}tiffness
  effects on behavior of bulk material},}\ }\href
  {https://doi.org/https://doi.org/10.1016/j.partic.2013.03.006} {\bibfield
  {journal} {\bibinfo  {journal} {Particuology}\ }\textbf {\bibinfo {volume}
  {12}},\ \bibinfo {pages} {107--112} (\bibinfo {year} {2014})}\BibitemShut
  {NoStop}%
\bibitem [{\citenamefont {Gloss}(2000)}]{Gloss}%
  \BibitemOpen
  \bibfield  {author} {\bibinfo {author} {\bibfnamefont {K.~T.}\ \bibnamefont
  {Gloss}},\ }\emph {\bibinfo {title} {A Photoelastic Investigation Into The
  Effects Of Cracks And Boundary Conditions On Stress Intensity Factors In
  Bonded Specimens}},\ \href@noop {} {Ph.D. thesis},\ \bibinfo  {school}
  {Virginia Tech} (\bibinfo {year} {2000})\BibitemShut {NoStop}%
\bibitem [{\citenamefont {Derakhshani}, \citenamefont {Schott},\ and\
  \citenamefont {Lodewijks}(2015)}]{Derakhshani}%
  \BibitemOpen
  \bibfield  {author} {\bibinfo {author} {\bibfnamefont {S.~M.}\ \bibnamefont
  {Derakhshani}}, \bibinfo {author} {\bibfnamefont {D.~L.}\ \bibnamefont
  {Schott}},\ and\ \bibinfo {author} {\bibfnamefont {G.}~\bibnamefont
  {Lodewijks}},\ }\bibfield  {title} {\enquote {\bibinfo {title} {Micro–macro
  properties of quartz sand: {E}xperimental investigation and {DEM}
  simulation},}\ }\href
  {https://doi.org/https://doi.org/10.1016/j.powtec.2014.08.072} {\bibfield
  {journal} {\bibinfo  {journal} {Powder Technol.}\ }\textbf {\bibinfo {volume}
  {269}},\ \bibinfo {pages} {127--138} (\bibinfo {year} {2015})}\BibitemShut
  {NoStop}%
\bibitem [{\citenamefont {Carvalho}\ and\ \citenamefont
  {Franklin}(2022)}]{Supplemental2}%
  \BibitemOpen
  \bibfield  {author} {\bibinfo {author} {\bibfnamefont {D.~D.}\ \bibnamefont
  {Carvalho}}\ and\ \bibinfo {author} {\bibfnamefont {E.~M.}\ \bibnamefont
  {Franklin}},\ }\href@noop {} {\bibfield  {journal} {\bibinfo  {journal}
  {{LIGGGHTS} input and output files, a code implemented in {LIGGGHTS} to
  introduce basal friction effects, and Octave scripts for post-processing the
  outputs are available on Mendeley Data,
  https://data.mendeley.com/datasets/39wgn3jtxb}\ } (\bibinfo {year}
  {2022})}\BibitemShut {NoStop}%
\end{thebibliography}%

\end{document}